\documentclass[1p,review,11pt]{elsarticle}

\usepackage{fancyhdr} 
\usepackage[american]{babel}
\usepackage[latin1]{inputenc}
\usepackage[T1]{fontenc}
\usepackage{bm}
\usepackage{amsmath}
\usepackage{caption}
\usepackage{lscape}
\usepackage{latexsym}
\usepackage{amssymb}
\usepackage{amsfonts}
\usepackage{multirow}
\usepackage{setspace}
\usepackage{paralist}
\usepackage[percent]{overpic}
\usepackage{framed}
\usepackage{multicol}

\usepackage{booktabs}
\usepackage{adjustbox}
\usepackage{makecell}
\usepackage{hyperref}
\usepackage{cleveref}
\usepackage{graphicx}
\usepackage{subcaption}
\usepackage[table]{xcolor}
\usepackage{arydshln}
\biboptions{sort&compress}

\immediate\write18{%
makeindex -s nomencl.ist -o \jobname.nls -t \jobname.nlg \jobname.nlo%
}
\usepackage{nomencl}
\makenomenclature
\renewcommand*\nompreamble{\begin{multicols}{2}}
\renewcommand*\nompostamble{\end{multicols}}

{ \begin{list}%
	{$\Box$}%
	{\setlength{\labelwidth}{30pt}%
	 \setlength{\leftmargin}{35pt}}}%
{ \end{list} }

\newcommand{\bu}{\bm{u}}

\newcommand{\br}{\bm{r}}

\newcommand{\bF}{\bm{F}}

\newcommand{\bT}{\bm{T}}
\newcommand{\bw}{\bm{w}}
\newcommand{\bb}{\bm{b}}
\newcommand{\bRe}{\bm{\mathrm{\Re}}}

\def\L{\mathcal{L}}    

\def\Re{{\mathcal{R}e}} 

\DeclareGraphicsExtensions{.pdf,.jpg,.png,.PNG}

\journal{International Journal of Multiphase Flow}

\begin{document}
	
\begin{frontmatter}

\title{{\bf Physics-informed neural network for modelling force and torque fluctuations in a random array of bidisperse spheres}}

\vspace*{-8mm}
\author[UBCME]{Zihao Cheng}
\author[UBCM,UBCCH]{Anthony Wachs\corref{cor}}
\ead{wachs@math.ubc.ca}
\cortext[cor]{Corresponding author.}
\address[UBCM]{Department of Mathematics, University of British Columbia, Vancouver, BC, Canada.}
\address[UBCME]{Department of Mechanical Engineering, University of British Columbia, Vancouver, BC, Canada.}
\address[UBCCH]{Department of Chemical \& Biological Engineering, University of British Columbia, Vancouver, BC, Canada.}

\vspace*{-8mm}

\begin{abstract}
We present a physics-informed neural network (PINN) model to predict the hydrodynamic force and torque fluctuations in a random array of stationary bidisperse spheres. The PINN model is formulated based on two hypotheses: (\expandafter{\romannumeral1}) pairwise interaction assumption that approximates the total force/torque exerted on a target sphere by linear superposition of individual contributions from a finite number of influential neighbors; (\expandafter{\romannumeral1}) unified function representation that suggests a single functional form to describe the contribution from different neighbors based on the observation of probability distribution maps obtained with various binary interaction modes in bidisperse particle-laden flows. On this basis, we accordingly establish a compact PINN architecture to evaluate individual force/torque contribution of influential neighbors through the same neural network block which tremendously reduces the number of unknown parameters, and ultimately compute the total force/torque exerted on target sphere by their weighted sum. We compare the model predictions to Particle-Resolved Direct Numerical Simulation (PR-DNS) data of eight different cases in a range of Reynolds number $1\leq\Re\leq100$, solid volume fraction $10\%\leq\phi\leq40\%$, sphere diameter ratio $1.5\leq d_l^*/d_s^*\leq2.5$ and volume ratio $1\leq V_l^*/V_s^*\leq 4$, which demonstrates excellent performance with an optimal $R^2\approx0.9$ for both force and torque predictions. We establish a universal model that is applicable within the aforementioned input space, and examine its interpolation capability to the unseen data with multiple additional datasets. Finally, we extract and illustrate the interpretable information of our PINN model in binary and trinary interactions, and discuss its potential extensions to other particle-laden flow problems with more complicated scenarios and Eulerian-Lagrangian simulations as a superior alternative to the classic average drag laws.\newline
\end{abstract}


\end{frontmatter}



\section{Introduction \label{sec-intro}}

A great portion of numerical efforts in the multiphase flow community over the last three decades has been dedicated to particle-laden flows that are commonly encountered in nature (river bank erosion, volcanic eruption, etc.) and industry (fluid catalytic cracking, coal combustion, etc.). Such a dispersed multiphase flow typically involves a massive amount of particles with distinct properties immersed in a homogeneous fluid. The system scale is often much larger than the characteristic length of the comparatively tiny particles. With the rapid development of computational methods in fluid mechanics, we are now able to complement Eulerian-Eulerian (EE) and Eulerian-Lagrangian (EL) simulations where the boundary layers are unresolved at particle length scales (even though the particles can be individually tracked in EL simulations) with high fidelity Particle-Resolved Direct Numerical Simulation (PR-DNS) that is capable of accessing the hydrodynamic force/torque exerted on each individual particle and hence of providing abundant microscale details to reflect the non-negligible force/torque fluctuations induced by the local microstructure that are not accessible with non-fully resolved approaches \cite{Hoef2008,Balachandar2009,Tenneti2014,Brandt2022}.

Even though we have been equipped with several versatile and robust numerical methods for large-scale computations and benefitted from the expeditious advance in computing power, PR-DNS of such a complicated and non-linear particle-laden flow problem is nevertheless expensive in terms of computing time and resource consumption. Therefore, to provide a theoretical insight into the neighbor-induced disturbance in particle-laden flows and avoid repeating the intrinsically same simulation of any given flow configuration, particular emphasis has been placed on developing a micro/mesoscale model to account for the force/torque fluctuations. Some early efforts attempted to apply modifications based on the classical average drag closures to incorporate the effect of local microstructure heterogeneity with different factors, e.g. standard deviation of solid concentration fluctuation \cite{Wang2011}, local drift velocity \cite{Rubinstein2017}, gradient of solid concentration \cite{Su2017} and its angle with the flow direction \cite{Li2017}. In recent years, a so-called stochastic approach drew much attention and several formulations of drag distributions have been proposed based on the statistical analysis of PR-DNS results, with each claiming satisfactory agreement compared to the simulation data obtained in freely evolving liquid-solid and gas-solid suspensions \cite{Tenneti2016,Esteghamatian2018,Lattanzi2020,Lattanzi2022,Hardy2022}. With a distinct perspective, Akiki et al. proposed a pairwise interaction extended point-particle (PIEP) model that decomposes the microscale flow around a target/reference particle into the perturbation flows induced by its neighbors, and the influence due to each neighbor can be separately evaluated and linearly superposed to obtain the total force exerted on the target particle. A similar idea was presented in a recent work of Seyed-Ahmadi and Wachs \cite{SeyedAhmadi2020}, in which a microstructure-informed probability-driven point-particle (MPP) model was established to predict the force/torque fluctuations based on conditioned probability distributions.

Regardless of the categories that distinguish different micro/mesoscale models for force/torque fluctuations in particle-laden flows, they are all deduced from and/or compared with PR-DNS data. Thanks to the collective contribution of the entire CFD community, we have been accumulating an expanding database obtained from PR-DNS of many particle-laden flow configurations. Considering the growing demand for processing big data in our problems of interest and the unprecedented development of data science, we come forward naturally to seek the help of machine learning (ML) techniques to tackle the still unresolved problem of accurate force/torque point-particle models. The utilization of ML concepts in fluid mechanics can even be dated back to the early 1940s for turbulence study \cite{KOLMOGOROV1941}. We refer the interested reader to the review paper by Burton et al. \cite{Brunton2020} and the literature therein for a comprehensive understanding of various branches of ML-related works in flow mechanics. Among the extensive applications, physics-informed learning, especially the family of physics-informed neural networks (PINN) pioneered by Raissi et al. \cite{Raissi2017a,Raissi2018,Raissi2019a,Raissi2019} has stirred a lively discussion in the scientific computing community. The conventional ML approaches that leverage a purely data-driven fitting process may not be able to understand or exploit the interpretable information and physical patterns underlying the data deluge, thus leading to inconsistent predictions or even poor performance. On the contrary, the physics-informed learning machines take into account the fundamental physics that governs the observation data, e.g., symmetries, invariance and conservation principles, and are hence capable of fulfilling robust generalization performance even with sparse and noisy data of complex physical systems. Therefore, an increasing number of applications are inspired to incorporate the existing knowledge and constraints to the corresponding learning machines for different real-world problems \cite{Cai2021,Karniadakis2021,Cuomo2022,Zhu2022}. Apart from the original PINN model proposed by Raissi et al. which successfully reproduced the hydrodynamics of the flow past a fixed cylinder and vortex-induced vibrations in the case of a moving cylinder \cite{Raissi2019,Raissi2019a}, numerous PINN variants have been developed to increase the overall accuracy and efficiency to solve the Navier-Stokes equations, e.g., cPINN with domain decomposition technique \cite{Jagtap2020}, hp-VPINN based on the sub-domain Petrov-Galekin method \cite{Kharazmi2021}, can-PINN with coupled automatic differentiation (AD) and numerical differentiation (ND) methods \cite{Chiu2022} and NSFnets \cite{Jin2021}. The interested reader is referred to \cite{Faroughi2022} for a non-exhaustive list of various PINN architectures to model assorted flow configurations.

Back to our problem of particle-laden flows, the mission is not to find an alternative to the conventional CFD methods to solve the forward and inverse problems involving various partial differential equations, but to utilize a physics-informed data-driven approach to find the statistical relationship between hydrodynamic force/torque and different independent variables throughout innumerable simulation results, and presumably substitute this relationship in the form of an advanced model into EL simulations with satisfactory accuracy and efficiency. This has been an active research topic in the last five years, about which we can find a variety of related works in the literature, e.g., drag predictions of 2D \cite{Viquerat2020} and 3D non-spherical particles \cite{Xin2022} in both compressible and incompressible flows, rough surfaces\cite{Lee2022} and translational particles in viscoelastic fluids \cite{Faroughi2022a}. Several attempts of leveraging the power of ML have been devoted to multi-particle systems as well. He and Tafti \cite{He2019} considered the relative positions of up to 15 nearest neighbors together with the Reynolds number $\Re$ and the solid volume fraction $\phi$, which are the only independent variables in classic average drag laws, to improve the drag predictions in dense fluid-particle flows. Muralidhar et al. \cite{Muralidhar2019,Muralidhar2020} introduced a PhyDNN model that demonstrated a better performance by learning with several physical intermediate variables under the same consideration of 15 nearest neighbors as inputs. Moore et al. \cite{Moore2019} extended the physical PIEP model to a hybrid version that combines it with a data-driven approach to improve the model performance at higher $\phi$. Seyed-Ahmadi and Wachs \cite{SeyedAhmadi2022} proposed a new PINN architecture in which the NN block (of fully connected layers) was shared by all the influential neighbors based on a unified function representation, thus greatly reducing the number of model parameters and enabling to take more neighbors into account without changing the PINN framework. Siddani and Balachandar \cite{Siddani2023} presented an equivariant NN and trained their universal model based on datasets from two sources in a range of $0.25\leq\Re\leq250$ and $0\leq\phi\leq0.4$. Furthermore, there exists a few PINN models for predicting forces exerted on non-spherical particles in dense particle-laden flows, e.g., ellipsoidal particles in 2D \cite{Shirzadi2023} and 3D \cite{Ashwin2022}.

Even though we are embracing a new era of physics-informed ML with an ever-increasing number of its applications in the fluid mechanics field, including particle-laden flow problems that are of particular interest to us, it is notable that the vast majority of those studies have exclusively focused on  flow configurations with monodisperse particles, i.e., particles of the same size. To the best of our knowledge, only Hwang et al. \cite{Hwang2023} presented a deep learning-based approach to model the drag force in polydisperse particle-laden flows, but only a dilute system with $\phi=0.5\%$ was considered in that work. Polydisperse particle-laden flows are not uncommon but rather highly prevalent in the real world, and of great research interest for their rich hydrodynamics. The existing gap in the extension of ML techniques to dense polydisperse systems may be attributed to the data paucity as PR-DNS of such complicated flows requires more computing resources in order to properly resolve the smallest particles. In fact, without the comprehensive understanding of force/torque distributions in each and every particle class (particles in each class have the same size), it is impossible to provide insightful knowledge to learning machines and optimize the architectures accordingly. As a preliminary study, we conducted a series of PR-DNS of the flow past a random array of stationary bidisperse and polydisperse spheres to investigate the microstructure-induced force/torque fluctuations in these flow configurations \cite{Cheng2023}. The simulation results indicated that every component of the hydrodynamic force and torque exerted on each particle class follows a Gaussian distribution. We also confirmed that different binary interaction modes in bidisperse particle-laden flows contain the same qualitative characteristics based on the comparison of corresponding probability distribution maps, which ultimately determines the framework of our PINN model. Consequently, we select the flow past stationary bidisperse spheres as a starting point of our "proof-of-concept" study. In this work, we establish a  PINN model similar to that proposed by Seyed-Ahmadi and Wachs \cite{SeyedAhmadi2022} in terms of common NN block for all influential neighbors, and account for the effect of both the local flow and particle microstructure through an element-wise multiplication operation. The individual contribution of each neighbor is evaluated separately but processed by the same NN block according to the pairwise interaction assumption and unified function representation, and linearly superposed to acquire the total force/torque exerted on a target sphere. Apart from the great accuracy and generalization performance of the prediction of force/torque fluctuations in bidisperse particle-laden flows, we can also see the potential of extending such a compact yet robust PINN model to more complicated polydisperse spheres to later plug it in EL simulations.

The rest of paper is organized as follows. In Section \ref{sec-data}, we describe the flow configurations and key parameters of the selected PR-DNS cases, followed by the dataset preparation procedures. In Section \ref{sec-network}, we start with the model formulation based on the physical fundamentals provided by the previous research and our preliminary numerical findings, then transition to the PINN architecture used in the present work and corresponding hyper-parameters tuning process. In Section \ref{sec-results}, we demonstrate the performance of two variants of the model for force/torque predictions, the former variant corresponding to a PINN model per set of $\left(\Re,\phi,d_l^*/d_s^*,V_l^*/V_s^*\right)$ values and the later variant corresponding to a universal model trained with all available datasets. We evaluate the interpolation capability of the universal model, and explore the interpretability of our PINN models in terms of binary and trinary interactions. Finally, we share our outlook for future research in Section \ref{sec-con}.


\section{Dataset construction \label{sec-data}}

\subsection{Numerical method and PR-DNS cases} \label{sec-data-dns}

To provide a sufficiently large and reliable database for the subsequent neural network training procedures, we implement a series of PR-DNS of the flow past a random array of stationary bidisperse spheres with an Immersed Boundary-lattice Boltzmann (IB-LB) method that was described comprehensively in our prior work \cite{Cheng2022} (we refer the interested reader to this work for details). We implemented the solver on the open-source platform {\it Basilisk} \cite{Popinet2015} to enable MPI parallel computing on an adaptive octree-grid, which facilitates the current large-scale simulations on up to 384 processors with satisfactory scalability. Apart from the high efficiency, our previous numerical results demonstrate very satisfactory accuracy of the simulations of the flow past a random array of moderately to strongly bidisperse spheres as well as of strongly polydisperse spheres. Relying on the same framework, we extend these bidisperse array simulations and the flow system can be characterized as follows.

We consider a pressure-driven flow with average velocity $\bm{u_0^*}$ (superscript $\dagger^*$ describes dimensional quantities) around a stationary sphere assembly of two groups with different diameter $d^*$ in a tri-periodic cubic box with edge length $L^*$. The solid volume fraction $\phi$ of the system can be corresponding expressed as:
\begin{equation}
	\phi = \frac{V_s^*+V_l^*}{\left(L^*\right)^3} = \frac{\pi \left[N_s\left(d_s^*\right)^3+N_l\left(d_l^*\right)^3\right]}{6\left(L^*\right)^3}
\end{equation}  
where the subscripts $\dagger_s$ and $\dagger_l$ represent the small and large spheres respectively, and $V^*$ and $N$ denote the total volume and number of spheres in each group. Here we introduce the most commonly used characteristic length for bidisperse spheres, the Sauter mean diameter:
\begin{equation}
	\langle d^* \rangle = \frac{N_s\left(d_s^*\right)^3+N_l\left(d_l^*\right)^3}{N_s\left(d_s^*\right)^2+N_l\left(d_l^*\right)^2}
\end{equation}
which leads us to define the particle Reynolds number $\Re$ as follows:
\begin{equation}
	\Re = \frac{\left(1-\phi\right)\left|\bm{u_0^*}\right|\langle d^* \rangle}{\nu^*}=\frac{\left|\bm{U^*}\right|\langle d^* \rangle}{\nu^*}
\end{equation}
where $\nu^*$ is the fluid kinematic viscosity and $\bm{U^*}=\left(1-\phi\right)\bm{u_0^*}$ is the superficial velocity. We constantly impose the flow along the $x$-direction in this work, and record the steady state data after a sufficiently long simulation time.

\begin{table}
	\centering
	\renewcommand{\arraystretch}{1.2}
	\begin{tabular}{ccccccc} 
		\toprule
		$\Re$ & $\phi$ & $d_l^*/d_s^*$ & $N_l$ & $N_s$ & $V_l^*/V_s^*$ & NOR \\
		\midrule
		1 & 15\% & 2 & 192 & 768 & 2 & 20 \\
		10 & 15\% & 2 & 192 & 768 & 2 & 20 \\
		100 & 15\% & 2 & 192 & 768 & 2 & 20 \\
		1 & 30\% & 2 & 384 & 1536 & 2 & 10 \\
		10 & 30\% & 2 & 384 & 1536 & 2 & 10 \\
		100 & 30\% & 2 & 384 & 1536 & 2 & 10 \\
		10 & 20\% & 1.5 & 453 & 1528 & 1 & 10 \\
		10 & 40\% & 2.5 & 313 & 1222 & 4 & 13 \\
		\bottomrule
	\end{tabular}
	\caption{Parameters of PR-DNS cases for generating datasets in the present study. NOR: number of realizations.}
	\label{prams_tab}
\end{table}

Table \ref{prams_tab} lists the eight PR-DNS cases selected in this work, which cover $1\leq\Re\leq100$, $15\%\leq\phi\leq40\%$, diameter and volume ratios of bidisperse spheres $1.5\leq d_l^*/d_s^*\leq2.5$ and $1\leq V_l^*/V_s^*\leq 4$, respectively. We design each and every case with $L^*=20d_s^*$ and seed the flow domain with $O(10^3)$ spheres. We perform $10\sim20$ realizations to generate approximately 20000 data points (among which around 4000 are related to large spheres) for each case.

\subsection{Data post-processing} \label{sec-data-proc}

We organize the dataset of each simulation case in the following manner: we find the $P$ ($P=100$ in this work) nearest neighbors of each sphere and sort them by their center-to-center distance to the target sphere. The tri-periodic boundary conditions have been considered during the search procedure. Each neighbor features four quantities: its diameter and three components of inter-center distance. Following the neighbor information, we attach six key parameters of interest to each target sphere $i$: three components of both the total fluid-particle interaction force $\bF_i^*$ and torque $\bT_i^*$. Furthermore, we include the target sphere diameter $d_i^*$, the local volume-averaged fluid velocity $\overline{\bu}_i^*$ and the local solid volume fraction $\phi_i$ as additional information to depict the corresponding local microstructure of each target sphere as suggested in \cite{SeyedAhmadi2022,Siddani2023}. In this work, we consider as "{\it local}" a cubic box centered at each target sphere with an edge length of $5d_s^*$ and $6d_s^*$ for small and large spheres respectively with $\phi\leq20\%$, and $4d_s^*$ and $5d_s^*$ otherwise.

Now we collect the data for all spheres and adopt the following scaling factors for the dimensional quantities:
\begin{itemize}
	\item length scale: $d_s^*$
	\item velocity scale: $\left|\bm{U^*}\right|$
	\item force scale: $3\pi\mu^* d_i^*\left|\bm{\overline{\bu}_i^*}\right|$, similar to the Stokes drag, where $\mu^*$ is the fluid dynamic viscosity.
	\item torque scale: $\mu^* \left(d_i^*\right)^2\left|\bm{\overline{\bu}_i^*}\right|$
\end{itemize}
After the aforementioned non-dimensionalization procedures, we constructed eight datasets corresponding to eight simulation cases. To characteristically distinguish them, we label each dataset in the following manner  $\left(\Re,\phi,d_l^*/d_s^*,V_l^*/V_s^*\right)$ hereafter (please note that since $d_s=d_s^*/d_s^*=1$, we can also use $d_l^*/d_s^*=d_l$ for conciseness, but prefer the former notation to emphasize the fact that the third parameter in the label of each dataset is indeed a diameter ratio). 

\begin{figure}
	\centering
	\captionsetup[subfigure]{oneside,margin={0.5cm,0.cm}}	
	\resizebox{0.7\linewidth}{!}{\includegraphics{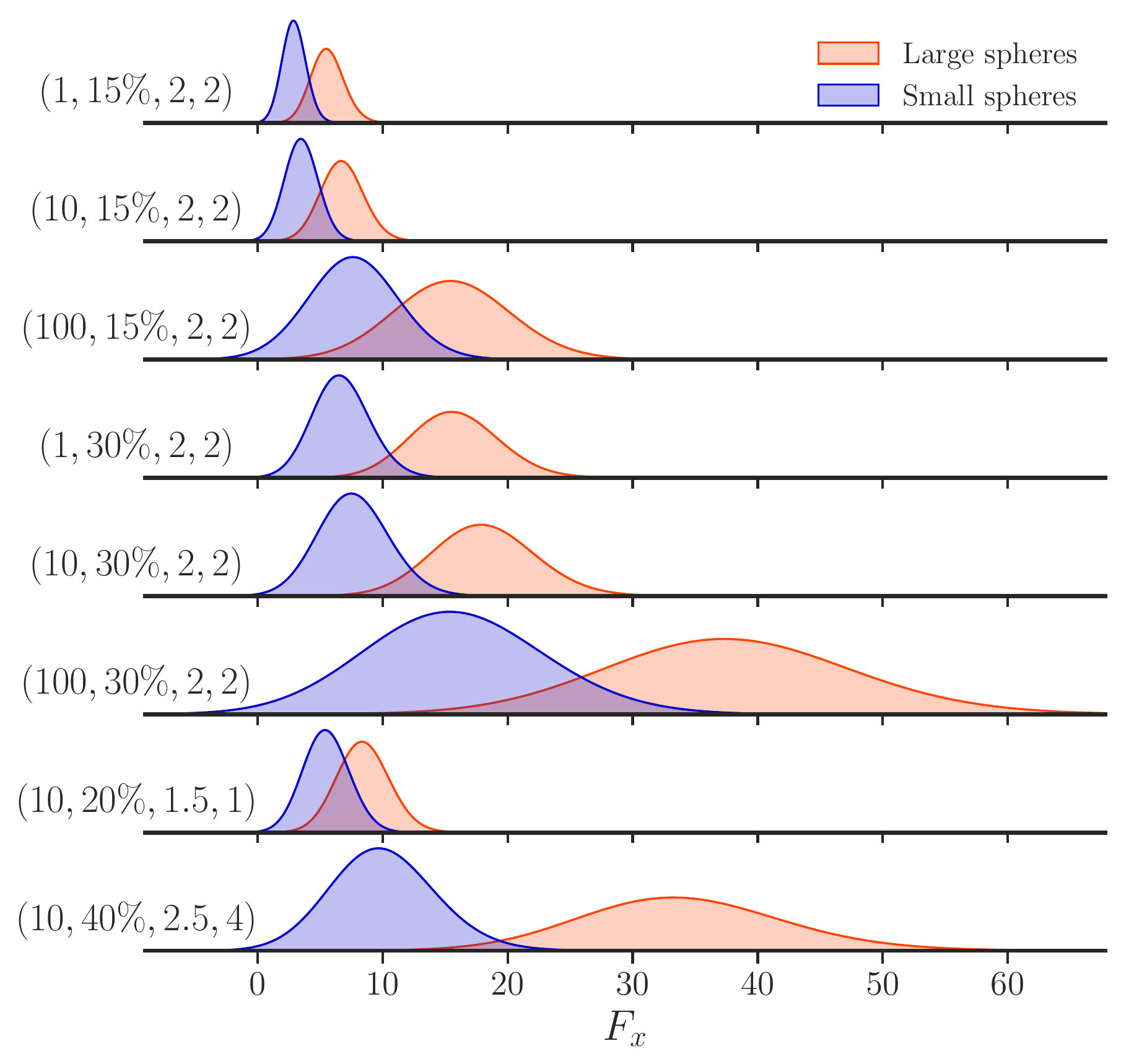}}
	\caption{Comparison of $F_x$ distribution of large and small spheres in each case obtained with PR-DNS.}
	\label{Fx-plot}
\end{figure}

Figure \ref{Fx-plot} compares the distribution of streamwise force $F_x$ exerted on bidisperse spheres in each computed case. Please note that $F_x$ is the only force component to have a non-zero mean value, while $F_y$, $F_z$ and all torque components have a zero mean value. We remind the reader that we use a different scaling for the $y$ axis in each case of Figure \ref{Fx-plot}. While the shaded areas are visually not equal in size, they are all equal to $1$ as expected. From the plot, we attempt to convey two messages: (\expandafter{\romannumeral1}) there exists a certain pattern of the force and torque distribution in bidisperse particle-laden flow problems (and also polydisperse cases), i.e., hydrodynamic force and torque exerted on each particle class both follow a Gaussian distribution, which constitute a foundation to successfully train a neural network; (\expandafter{\romannumeral2}) the distribution of each group exhibits different mean values, standard deviations and other statistical characteristics, and each case shows different extents of overlapping between two groups of data. These features collectively add complexity to our problem in comparison to the monodisperse particle-laden flows, especially when aiming to establish a universal model without the need to separate the small and large spheres.

\section{Neural network architecture and training process \label{sec-network}}

\subsection{Model formulation}

The force and torque distributions in a random array of stationary spheres are dependent on the unique local microstructure surrounding each sphere as well as the macroscopic fluid properties, hence of great difficulty to identify and predict. The pairwise interaction assumption, which decomposes the total force/torque exerted on a target sphere into the individual contribution from only a few influential neighbors, significantly simplifies the analysis of force and torque fluctuations in dense particle-laden flows, and paves the way for the development of several deterministic models, e.g. PIEP \cite{Akiki2017a,Akiki2017b,Moore2019,Balachandar2020}, MPP \cite{SeyedAhmadi2020}, PINN \cite{SeyedAhmadi2022} and equivariant NN models \cite{Siddani2023}, which all demonstrate good predictive performance in monodisperse particle-laden flow problems spanning a range from low to moderate $\Re$. In this work, we also formulate the model based on the pairwise interaction assumption and merely consider the binary interactions as a starting point.

Inspired by Seyed-Ahmadi and Wachs \cite{SeyedAhmadi2020,SeyedAhmadi2022}, the force $\bF_i$ and torque $\bT_i$ exerted on the target sphere $i$ can be approximated as:
\begin{subequations}\label{eq:piep}
	\begin{align}
		\bF_i & \approx \sum_{j=1}^{M}\Delta \bF_{j\rightarrow i} \approx \sum_{j=1}^{M}\alpha_jf_{j\rightarrow i}\left(\left\{\br_j,d_j\right\},\left\{\bRe_i,\phi_i,d_i\right\}\right)\label{eq:piep-a}\\
		\bT_i & \approx \sum_{j=1}^{M}\Delta \bT_{j\rightarrow i} \approx \sum_{j=1}^{M}\beta_jg_{j\rightarrow i}\left(\left\{\br_j,d_j\right\},\left\{\bRe_i,\phi_i,d_i\right\}\right)
	\end{align}
\end{subequations}
where $\Delta \bF_{j\rightarrow i}$ and $\Delta \bT_{j\rightarrow i}$ are the binary interaction force and torque induced by the $j$th neighbor with diameter $d_j$ and distance $\br_j$ to the target sphere $i$. $M$ denotes the number of influential neighbors considered in this work. $\bRe_i=d_i \overline{\bu}_i/\nu$ denotes the local $\Re$ of the mesoscale flow around target sphere $i$. It should be noted that $\Delta \bF_{j\rightarrow i}$ and $\Delta \bT_{j\rightarrow i}$ are typically interpreted as the perturbation influences of neighboring particles, thus for the streamwise force $F_x$ that has a non-zero mean value, an additional term due to the mesoscale flow should be added, i.e., $\bF_i \approx \langle \bF\rangle\left(\bRe_i,\phi_i,d_i\right) + \sum_{j=1}^{M}\Delta \bF_{j\rightarrow i}$. However, we remove the mesoscale contribution term in Equation \ref{eq:piep-a} and maintain a universal form for all components of both force and torque. In this regard, $\Delta \bF_{j\rightarrow i}$ and $\Delta \bT_{j\rightarrow i}$ can be interpreted as the contribution of $j$th neighbor through the "{\it local}" microstructure as well as the mesoscale flow, which has incorporated the collective effect of other spheres in this local volume in an average sense. Back to Equation \ref{eq:piep}, this means we evenly distribute the unary term $\langle \bF\rangle$ to each binary interaction term $\Delta\bF_{j\rightarrow i}$.

Our previous simulation results \cite{Cheng2023} provide us with another insight to further simplify this formulation: the probability distribution maps of different binary interaction modes (small-to-small, large-to-small, small-to-large and large-to-large spheres) exhibit identical qualitative features in terms of the critical regions of neighboring sphere that result into a certain condition on the force/torque exerted on the target sphere. Therefore, we are able to use a unified function representation for contributions from different neighbors as suggested by Seyed-Ahmadi and Wachs \cite{SeyedAhmadi2022} and consequently substitute the giant fully connected NN by a compact share-block NN (we will present this NN architecture in the next subsection). The unified functions for binary force and torque in this work, $f_{j\rightarrow i}$ and $g_{j\rightarrow i}$, are slightly more complicated than those in \cite{SeyedAhmadi2022} as they are not only dependent on the neighbor properties $\left\{\br_j,d_j\right\}$, but also on the mesoscale flow and local microstructure properties $\left\{\bRe_i,\phi_i,d_i\right\}$. This is also reflected in the design of our NN architecture.  As the final operation, the total force and torque exerted on a target sphere are evaluated through the weighted sums with $\alpha_j$ and $\beta_j$ based on the ordering of neighbors.

We would like to point out that we exclude the local volume ratio $V_{i,l}^*/V_{i,s}^*$ when describing the local microstructure while it would seem natural to include it. Actually, the global volume ratio $V_l^*/V_s^*$ was unexpectedly ignored in most average drag closures despite the fact that it has a non-negligible impact on the forces and torques exerted on bidispere spheres. To justify our choice of local quantities in the input space, we compared the predictions provided by models trained with and without $V_{i,l}^*/V_{i,s}^*$ as one input parameter, and it turned out that the model performance constantly reduces by a small amount, at most $2\%$, when taking $V_{i,l}^*/V_{i,s}^*$ into account. We consider this slight model performance reduction as insignificant, although it is partially unexpected. One plausible explanation is that the values of $V_{i,l}^*$ and $V_{i,s}^*$ can be calculated by $\phi_i$ and $\left\{\br_j,d_j\right\}$ of neighbors, thus $V_{i,l}^*/V_{i,s}^*$ introduces repeated information which eventually leads to predictions that are slightly less satisfactory. It is also possible that we  fully exploited the potential of our PINN models with the current low data density, and we can reasonably expect improved performance with $V_{i,l}^*/V_{i,s}^*$ as an input if we access to a substantially larger number of data points. Consequently, in the present work, we consider $\left\{\bRe_i,\phi_i,d_i\right\}$ only, instead of $\left\{\bRe_i,\phi_i,d_i,V_{i,l}^*/V_{i,s}^*\right\}$, to characterize the mesoscale flow and local microstructure around target sphere $i$.

Finally, we consider the same flow symmetry of pairwise interaction as introduced in \cite{Moore2019,SeyedAhmadi2022}. We refer our reader to these works and a graphical illustration can be found in Figure 1 of \cite{SeyedAhmadi2022}.

\subsection{Neural network architecture}

\begin{figure}
	\centering
	\captionsetup[subfigure]{oneside,margin={0.5cm,0.cm}}	
	\resizebox{0.95\linewidth}{!}{\includegraphics{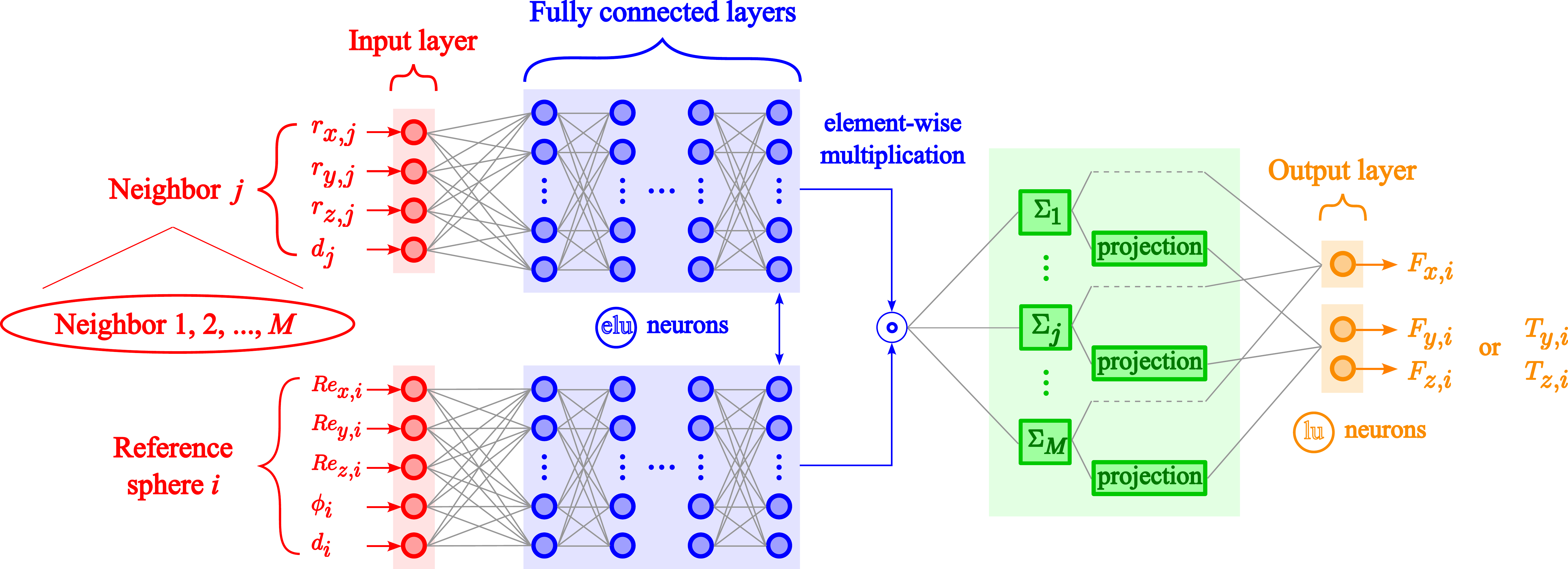}}
	\caption{Physics-informed neural network architecture used in the present work.}
	\label{model-plot}
\end{figure}

Figure \ref{model-plot} illustrates the NN architecture developed in this work. We can see in figure \ref{model-plot} that the network contains three main modules: an input layer, two blocks of fully connected layers and an output layer, each consisting of several neurons which are the fundamental units in a NN to process the input signals through a so-called activation function and assist the NN to learn the non-linearity after multiple-layer transformations. Inspired by Siddani and Balachandar \cite{Siddani2023}, we introduce an intermediate step of element-wise multiplication to incorporate the local flow effects, and a same projection step for the lateral components of force and torque as adopted by Seyed-Ahmadi and Wachs \cite{SeyedAhmadi2022}.

Now we elaborate on each module and intermediate step: for the input layer, we need the properties of both the target sphere and its $M$ influential neighbors, i.e. $\left\{\bRe_i,\phi_i,d_i\right\}$ and $\left\{\br_j,d_j\right\}$ respectively as mentioned above. Each subset of input parameters is then fed into an individual block of fully connected layers (also called hidden layers) with the same width, i.e., the same number of neurons in one layer. Please note that as suggested by the unified function representation assumption, information of different neighbors is passed through the same block at this stage, thus significantly reducing the width (and presumably the length, i.e., the number of layers) of fully connected layers and hence the unknown parameters. Here we adopt the exponential linear unit (elu) \cite{Clevert2015} as the activation function of fully connected layers which leads to fast and precise learning of NN and avoids the vanishing gradient problem during back-propagation. Next, the outputs generated by two individual blocks experience an element-wise multiplication step to combine the contribution from each neighbor with the effects of local flow and microstructure of a target sphere. According to the flow symmetry, each scalar value that corresponds to the unweighted contribution from each neighbor is multiplied by appropriate direction vectors and then projected on $y$ and $z$-axis to obtain the lateral force and torque. Eventually, all scalars are processed by a linear unit (lu) at the output layer to perform a weighted sum that yields the total force and torque on a target sphere.

It is also worth mentioning that due to the features of our problem, we select a special loss function, the Huber loss to optimize the values of unknown parameters (weights and bias in each neuron). In bidisperse particle-laden flows, the streamwise forces exerted on large spheres are significantly larger than those of small spheres, whereas the number of larger spheres is much less than its counterpart. Therefore, the data points corresponding to large spheres are prone to be treated as "outliers" during the regression process. Our NN may neglect these "outliers" with a simple MAE (mean average error) function, or be trapped by them and predict poorly for the small spheres with a conventional MSE (mean square error) function. The Huber loss balances these two functions in the following form:
\begin{equation}
	\L\left(\bw,\bb\right)=
	\begin{cases}
		\frac{1}{2}\left(\bF_{i,PINN}-\bF_{i,DNS}\right)^2 & \left|\bF_{i,PINN}-\bF_{i,DNS}\right|\leq\delta \\
		\delta\left(\left|\bF_{i,PINN}-\bF_{i,DNS}\right|-\frac{1}{2}\delta\right) & otherwise
	\end{cases}
\end{equation}
where $\bw$ and $\bb$ are the weight and bias matrices respectively. In the present work, $\delta$ equals to the standard deviation of the distribution of force/torque exerted on large spheres.

\subsection{Hyper-parameters and implementation}\label{subsec:hyper-pm}

We presented the NN architecture in the previous subsection, while we still need to determine the values of several hyper-parameters, that significantly affect the learning process and performance of our NN and need to be fine-tuned in advance. Before introducing the hyper-parameter tuning process in this work, we would like to briefly present two notions: (\expandafter{\romannumeral1}) $5$-fold cross-validation: we randomly shuffle the dataset first and then split it into 4 folds of training set and 1 fold of test set for regression and evaluation purposes respectively. Each fold of data is retained as test set in turn to evaluate the model performance on the rest folds after each round of training, and the ultimate performance is averaged over all 5 rounds. In our implementation, 20\% of the training set is additionally separated and used as validation set during the training process to validate the model generalization. (\expandafter{\romannumeral2}) coefficient of determination $R^2$: a metric to quantify the model performance, which typically has the following form:
\begin{equation}
	R^2=1-\frac{\sum_{i=1}^{N}\left(\bF_{i,PINN}-\bF_{i,DNS}\right)^2}{\sum_{i=1}^{N}\left(\bF_{i,DNS}-\langle\bF_{i,DNS}\rangle\right)^2}
	\label{R2-each}
\end{equation}
where $\langle\bF_{i,DNS}\rangle$ is the average force/torque computed by PR-DNS. Considering the problem of interest in this work, we use the following formula to evaluate the overall performance:
\begin{equation}
	R^2=1-\frac{\sum_{i=1}^{N_s}\left(\bF_{(s,i),PINN}-\bF_{(s,i),DNS}\right)^2+\sum_{i=1}^{N_l}\left(\bF_{(l,i),PINN}-\bF_{(l,i),DNS}\right)^2}{\sum_{i=1}^{N_s}\left(\bF_{(s,i),DNS}-\langle\bF_{(s,i),DNS}\rangle\right)^2+\sum_{i=1}^{N_l}\left(\bF_{(l,i),DNS}-\langle\bF_{(l,i),DNS}\rangle\right)^2}
	\label{R2-overall}
\end{equation}
where we separate the force/torque exerted on small and large spheres. Please note that without special indication, the $R^2$ scores reported hereafter are evaluated on test sets through 5-fold cross-validation.

\begin{figure}
	\centering
	\captionsetup[subfigure]{oneside,margin={0.5cm,0.cm}}	
	\resizebox{\linewidth}{!}{\includegraphics{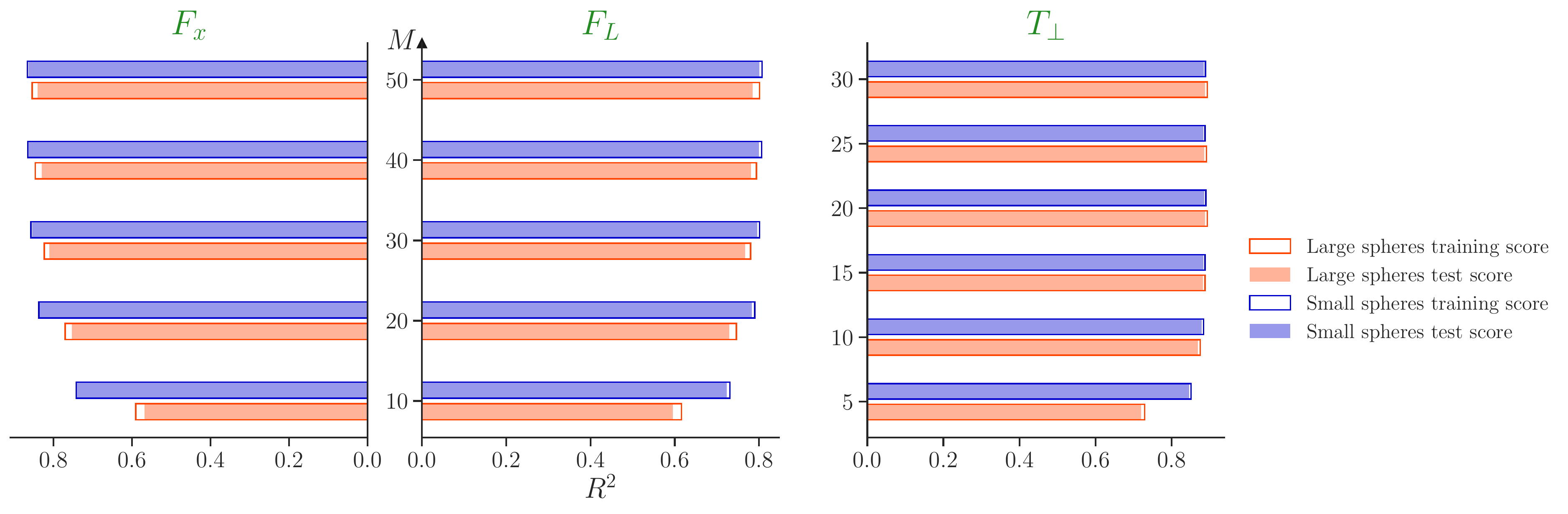}}
	\caption{$R^2$ scores of $F_x$, $F_L$ and $T_{\perp}$ for the increasing number of neighbors $M$ included in the PINN model. We select $M=50$ and $M=20$ for force and torque predictions respectively based on this convergence study.}
	\label{Ngb-conv-plot}
\end{figure}

We begin with the number of influential neighbors to be included in the input space. Neighbor-truncation error analysis of both \cite{SeyedAhmadi2022} and \cite{Siddani2023} shows that the $R^2$ converges with $25\sim30$ neighbors for force and 10 for torque predictions in monodisperse particle-laden flows. We carry out a similar convergence test for the streamwise force $F_x$, lateral force $F_L$ and lateral torque $T_{\perp}$ in bidisperse particle-laden flows and compare the training (hollow bar) and test (solid bar) $R^2$ scores for each component of both sphere classes with a different number of neighbors $M$ in Figure \ref{Ngb-conv-plot}. Once again, we gain two observations from the plot: (\expandafter{\romannumeral1}) the training and test $R^2$ are highly close to each other in every case regardless of $M$, demonstrating great generalization of our NN model, and (\expandafter{\romannumeral2}) $R^2$ reaches a plateau with $M=40\sim50$ for force and $15\sim20$ for torque, and consequently we select $M=50$ for the former and and $20$ for the latter.

\begin{figure}
	\centering
	\captionsetup[subfigure]{oneside,margin={0.5cm,0.cm}}	
	\resizebox{\linewidth}{!}{\includegraphics{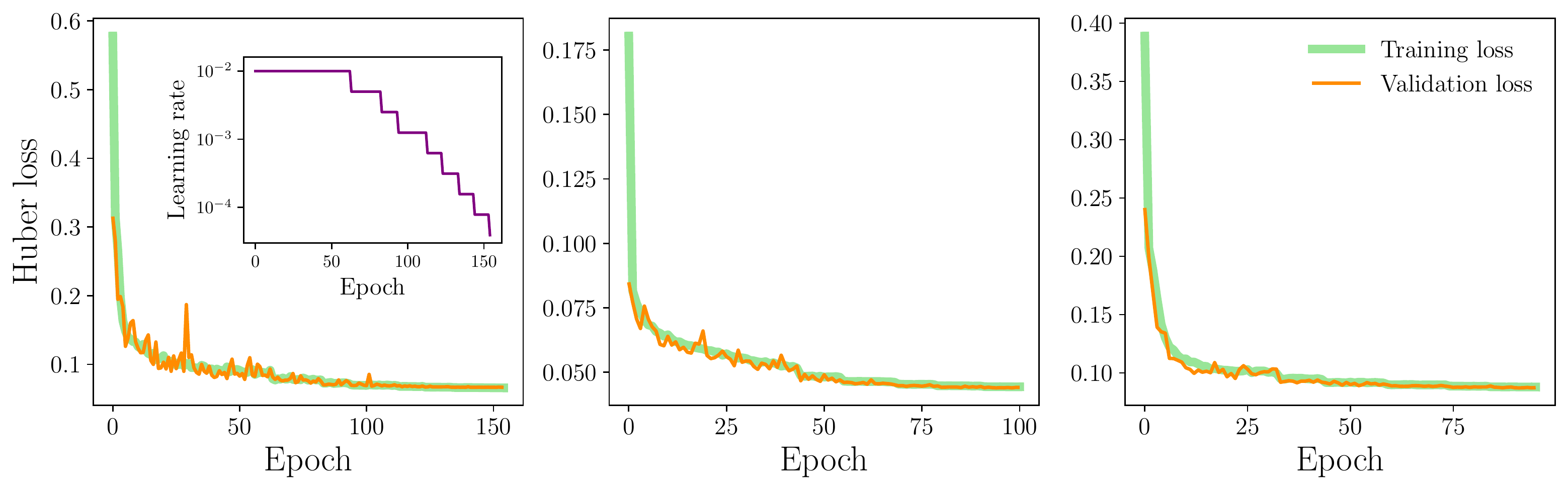}}
	\caption{Evolution of Huber loss during the model training process. From left to right: $F_x$, $F_L$ and $T_{\perp}$.}
	\label{huber-plot}
\end{figure}

We construct and implement our NN model using TensorFlow \cite{Abadi2016} with Keras API \cite{wood2022kerascv} in Python. The Nadam optimizer \cite{Dozat2016}, which is an extension to the Adam optimizer with Nesterov momentum, is adopted in our model training process. Two callbacks are applied to enhance the efficiency and accuracy: (\expandafter{\romannumeral1}) ReduceLROnPlateau: the learning rate will be reduced by a factor of 2 if no obvious improvement is achieved in terms of decreasing the total loss after 10 epochs; (\expandafter{\romannumeral2}) EarlyStopping: the training process will be terminated if no obvious improvement is achieved after 40 consecutive epochs. Figure \ref{huber-plot} exhibits the evolution of Huber losses of both the training and validation sets as well as the change in learning rate with increasing epochs. We can see that the training and validation losses both decrease rapidly at the early stage, and the adaptive learning rate assists the NN model to escape from the local minimum and continue to search for the global minimum. All of the loss functions reach the converged values and no over-fitting occurs during the training process. In addition, a batch size of 128 is used in all cases, and 3 layers of 30 neurons for force and 20 neurons for torque predictions are determined for the fully connected layer blocks after a grid-search within a broad range of hyper-parameters.

\section{Results and discussions \label{sec-results}}

\subsection{Performance over monodisperse cases}

To validate the performance of our PINN model, we start with the simulation cases of flow past a random array of stationary monodisperse spheres implemented by Seyed-Ahmadi and Wachs \cite{SeyedAhmadi2022}. These simulations cover broad ranges of $10\%\leq\phi\leq40\%$ and $0.2\leq\Re\leq150$. We perform the non-dimensionalization procedures described in Section \ref{sec-data-proc} to the original datasets. In our training process, $M=30$ and 10 are considered for force and torque predictions which are the same as those used in \cite{SeyedAhmadi2022}. Table \ref{mono_tab} compares the $R^2$ scores obtained with the PINN model proposed in \cite{SeyedAhmadi2022} and the present model for $F_x$, $F_L$ and $T_{\perp}$ (denoted $T_L$ in \cite{SeyedAhmadi2022}) respectively. We can see from the table that almost all $R^2$ have increased with the present model except the $R^2$ of $F_L$ in the case $\left(\Re,\phi\right)=\left(150,40\%\right)$, which reaches the upper limits of both parameters so that more data points should be expected to acquire a satisfactory prediction for the lateral force that varies more dramatically with higher $\Re$ and $\phi$. Unfortunately, we can only access approximately the same amount of data points for every case due to limited computing resources (in fact, the data points for the case $\left(150,40\%\right)$ are even fewer than those in the other cases \cite{SeyedAhmadi2022}). Nevertheless, our PINN model exhibits superior performance in all the other predictions, where more than half of them achieve a higher than 5\% increase in $R^2$. In addition, our results disclose a clear pattern that higher $R^2$ can be achieved at lower $\Re$. This improvement is physically understandable since the problem is less non-linear but not being observed in the previous results.

\begin{table}
	\centering
	\renewcommand{\arraystretch}{1.2}
	\begin{tabular}{cccccccc} 
		\toprule
		\multirow{2}{*}{$\phi$} & \multirow{2}{*}{$\Re$} & \multicolumn{2}{c}{$F_x$} & \multicolumn{2}{c}{$F_L$} & \multicolumn{2}{c}{$T_{\perp}$} \\
		\cmidrule(lr){3-4}\cmidrule(lr){5-6}\cmidrule(lr){7-8}
		~ & ~ & \cite{SeyedAhmadi2022} & Present & \cite{SeyedAhmadi2022} & Present & \cite{SeyedAhmadi2022} & Present \\
		\midrule
		\multirow{3}{*}{10\%} & 2 & 0.81 & 0.88 & 0.77 & 0.84 & 0.89 & 0.90 \\
		~ & 10 & 0.82 & 0.84 & 0.78 & 0.80 & 0.86 & 0.87 \\
		~ & 40 & 0.70 & 0.72 & 0.73 & 0.77 & 0.74 & 0.81 \\
		\midrule
		\multirow{4}{*}{20\%} & 0.2 & 0.68 & 0.79 & 0.57 & 0.78 & 0.74 & 0.89 \\
		~ & 2 & 0.73 & 0.78 & 0.68 & 0.76 & 0.88 & 0.90 \\
		~ & 40 & 0.71 & 0.74 & 0.67 & 0.72 & 0.79 & 0.81 \\
		~ & 150 & 0.61 & 0.63 & 0.56 & 0.65 & 0.56 & 0.71 \\
		\midrule
		\multirow{3}{*}{40\%} & 2 & 0.61 & 0.79 & 0.51 & 0.61 & 0.72 & 0.77 \\
		~ & 40 & 0.66 & 0.75 & 0.58 & 0.59 & 0.69 & 0.73 \\
		~ & 150 & 0.52 & 0.69 & 0.53 & 0.51 & 0.59 & 0.62 \\
		\bottomrule
	\end{tabular}
	\caption{$R^2$ score comparison of the PINN model proposed in \cite{SeyedAhmadi2022} and the present model for monodisperse particle-laden flows.}
	\label{mono_tab}
\end{table}

\subsection{Performance over bidisperse cases}

\begin{figure}
	\centering
	\captionsetup[subfigure]{oneside,margin={0.5cm,0.cm}}
	\begin{subfigure}[c]{0.95\linewidth}		
		\resizebox{\linewidth}{!}{\includegraphics{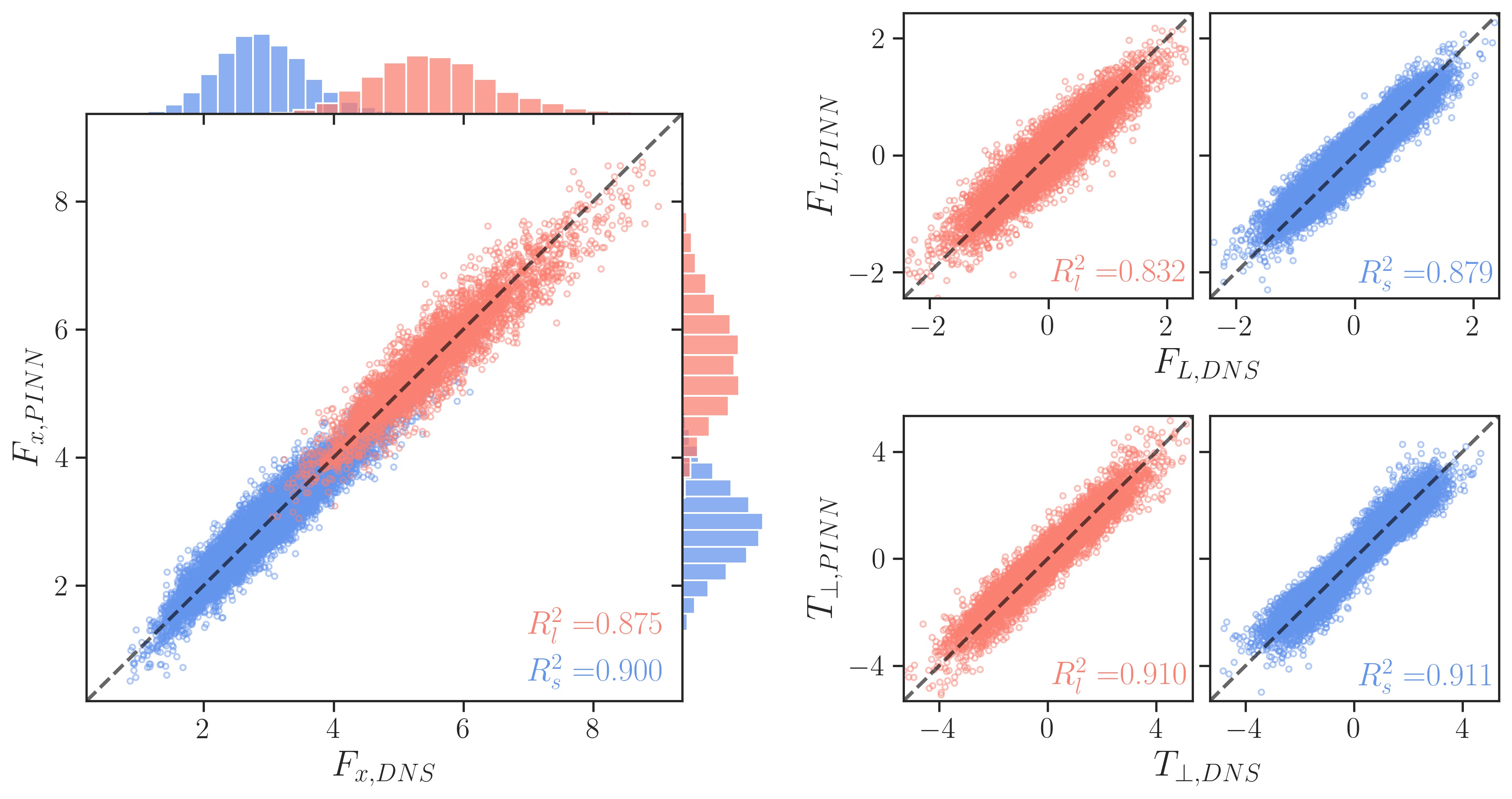}}
		\vspace{-0.7cm}
		\subcaption{$\left(\Re,\phi,d_l^*/d_s^*,V_l^*/V_s^*\right)=\left(1,15\%,2,2\right)$}
		\vspace{0.5cm}
		\label{corr_plot_Re1}
	\end{subfigure}
	\begin{subfigure}[c]{0.95\linewidth}		
		\resizebox{\linewidth}{!}{\includegraphics{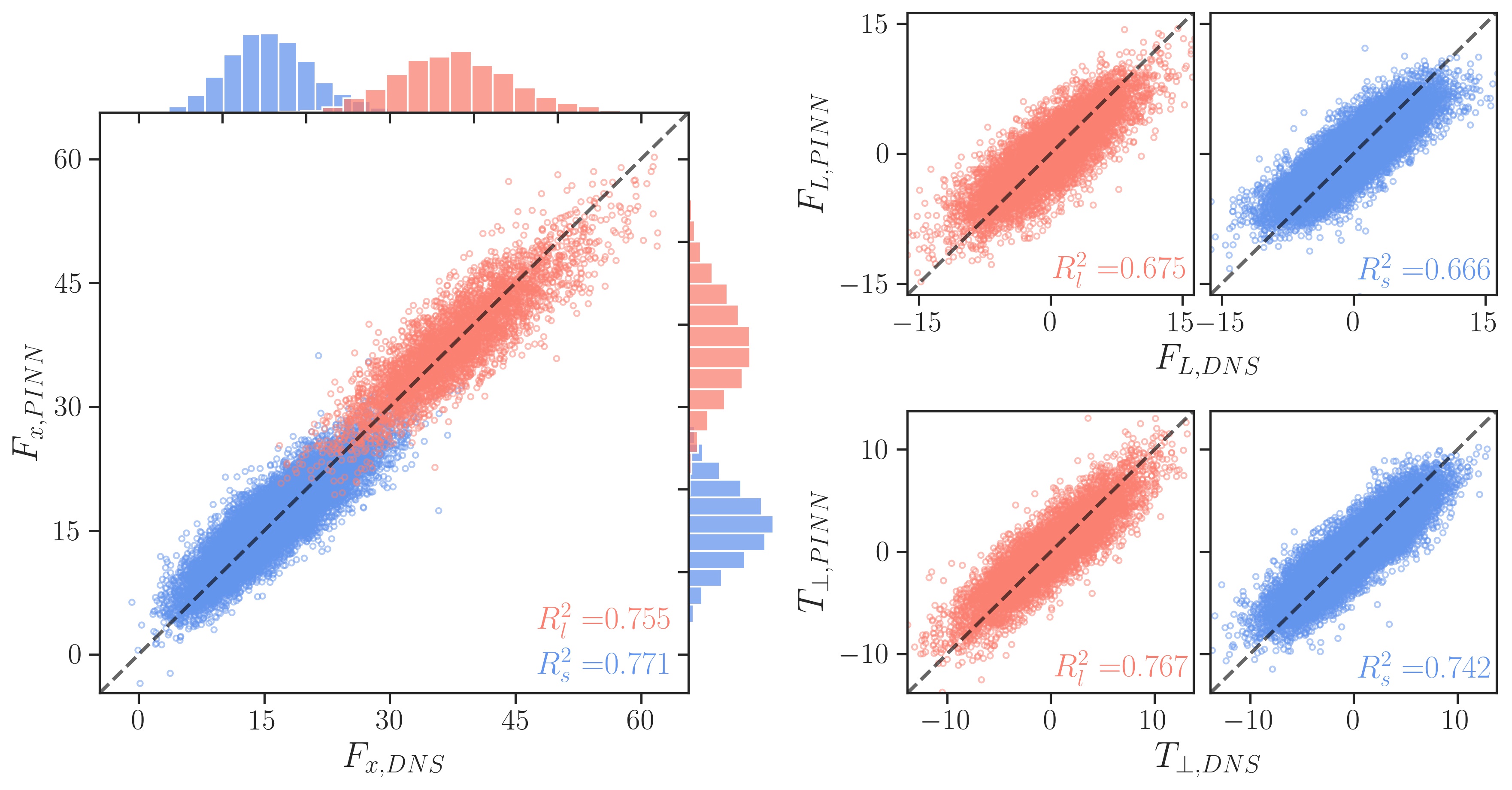}}
		\vspace{-0.7cm}
		\subcaption{$\left(\Re,\phi,d_l^*/d_s^*,V_l^*/V_s^*\right)=\left(100,30\%,2,2\right)$}
		\label{corr_plot_Re100}
	\end{subfigure}
	\caption{Comparison of $F_x$, $F_L$ and $T_{\perp}$ obtained with our PR-DNS and PINN predictions.}
	\label{corr_plot}
\end{figure}

Now we turn our attention to our problem of interest: flow past a random array of stationary bidisperse spheres described in Section \ref{sec-data-dns}. Figure \ref{corr_plot} shows the correlation plot of $F_x$, $F_L$ and $T_{\perp}$ exerted on all spheres obtained with our PR-DNS and the PINN model predictions in the two cases $\left(\Re,\phi,d_l^*/d_s^*,V_l^*/V_s^*\right)=\left(1,15\%,2,2\right)$ and $\left(100,30\%,2,2\right)$. These two cases correspond to the lower and upper limits of the two key parameters $\Re$ and $\phi$ where we indeed observe the best and worst model performance. To distinguish two sphere classes, we plot the large sphere data in red and the small sphere data in blue. On the $F_x$ plot where the large and small spheres have non-zero mean values, we attach the probability distributions of both groups aside for further comparison. The quantitative results, $R^2$ scores of each group obtained with Equation \ref{R2-each}, are also added to the corresponding subplot. We observe that both the red and blue scatters are located closely around the diagonal dashed line that represents the perfect prediction (i.e., $R^2=1$), which reflects that the error distribution is not tremendously skewed to one side and our PINN model gives ideal predictions in most cases. We also note that the data bands are much narrower at low $\Re$ and low $\phi$ than at large $\Re$ and large $\phi$. The probability distributions of $F_x$ obtained with PR-DNS and PINN model are remarkably similar in shape and range, and visually symmetric about the dashed line in both cases. Moreover, the converged shape at the two ends of each data band in all subplots indicates that our PINN model possesses good learning ability/generalization and is capable of providing acceptable predictions even in regions with sparse data. 

Table \ref{R2_tab} summarizes the individual and overall performance of $F_x$, $F_L$ and $T_{\perp}$ predictions in all eight cases. Several observations can be made from the comparison: (\expandafter{\romannumeral1}) The $R^2$ scores of the training and test sets are close to each other in most cases (difference $<$ 0.03), especially in the lateral torque predictions that exhibit the most satisfactory model generalization. (\expandafter{\romannumeral2}) The $R^2$ scores of small spheres are generally better than those of large spheres in force predictions, while they are highly comparable in torque predictions. The reason is that large spheres typically experience higher forces than their small counterparts, whereas they only possess a small proportion in the entire dataset, thus compromising the performance during the regression process. However, $T_{\perp}$ exerted on different sphere classes are quite close to each other, and less sensitive to the change of parameters $\Re$, $\phi$ and $d_l^*/d_s^*$. We refer the interested reader to our previous work for more details on simulation results of force and torque distributions in flows past a random array of stationary bidisperse spheres, and we will come back to this later in the statistical analysis. (\expandafter{\romannumeral3}) There exists the following clear pattern: if we increase any key parameter $\Re$, $\phi$ and $d_l^*/d_s^*$ while keeping the two other parameters constant, we will see a significant performance drop in all three predictions $F_x$, $F_L$ and $T_{\perp}$. (\expandafter{\romannumeral4}) The $T_{\perp}$ prediction provides the best overall performance even with a smaller number of influential neighbors than that used for force predictions, whereas $F_L$ produces the worst overall performance. This observation is in line with the results reported by Seyed-Ahmadi and Wachs \cite{SeyedAhmadi2022}. As a corollary, our PINN model generates great predictions in bidisperse particle-laden flows and the performance is either at the same level or even better than that obtained in  monodisperse particle-laden flows in \cite{SeyedAhmadi2022} and \cite{Siddani2023}. The overall $R^2$ spans $0.706\sim0.892$ for $F_x$, $0.669\sim0.871$ for $F_L$ and $0.741\sim0.910$ for $T_{\perp}$ predictions when $1\leq\Re\leq100$, $10\%\leq\phi\leq40\%$, $1.5\leq d_l^*/d_s^*\leq2.5$ and $1\leq V_l^*/V_s^*\leq 4$.

\begin{table}
	\centering
	\renewcommand{\arraystretch}{1.2}
	\begin{subtable}{\textwidth}
		\centering
		\begin{tabular}{cccccc} 
			\toprule
			\multirow{2}{*}{$\left(\Re,\phi,d_l^*/d_s^*,V_l^*/V_s^*\right)$} & \multicolumn{2}{c}{Large spheres $R^2$} & \multicolumn{2}{c}{Small spheres $R^2$} & 	\multirow{2}{*}{Overall $R^2$} \\
			\cmidrule(lr){2-3}\cmidrule(lr){4-5}
			~ & Training sets & Test sets & Training sets & Test sets & ~ \\
			\midrule
			$\left(1,15\%,2,2\right)$ & 0.898 & 0.875 & 0.911 & 0.900 & \textbf{0.892} \\
			$\left(10,15\%,2,2\right)$ & 0.888 & 0.852 & 0.892 & 0.874 & \textbf{0.868} \\
			$\left(100,15\%,2,2\right)$ & 0.777 & 0.712 & 0.743 & 0.704 & \textbf{0.706} \\
			$\left(1,30\%,2,2\right)$ & 0.893 & 0.849 & 0.901 & 0.884 & \textbf{0.871} \\
			$\left(10,30\%,2,2\right)$ & 0.888 & 0.838 & 0.893 & 0.871 & \textbf{0.859} \\
			$\left(100,30\%,2,2\right)$ & 0.818 & 0.755 & 0.800 & 0.771 & \textbf{0.766} \\
			$\left(10,20\%,1.5,1\right)$ & 0.909 & 0.858 & 0.894 & 0.869 & \textbf{0.864} \\
			$\left(10,40\%,2.5,4\right)$ & 0.852 & 0.829 & 0.876 & 0.860 & \textbf{0.852} \\
			\bottomrule
		\end{tabular}
		\caption{$F_x$}
		\label{R2_tab_Fx}
		\vspace{0.5cm}
	\end{subtable}
	\begin{subtable}{\textwidth}
		\centering
		\begin{tabular}{cccccc} 
			\toprule
			\multirow{2}{*}{$\left(\Re,\phi,d_l^*/d_s^*,V_l^*/V_s^*\right)$} & \multicolumn{2}{c}{Large spheres $R^2$} & \multicolumn{2}{c}{Small spheres $R^2$} & 	\multirow{2}{*}{Overall $R^2$} \\
			\cmidrule(lr){2-3}\cmidrule(lr){4-5}
			~ & Training sets & Test sets & Training sets & Test sets & ~ \\
			\midrule
			$\left(1,15\%,2,2\right)$ & 0.854 & 0.832 & 0.886 & 0.879 & \textbf{0.865} \\
			$\left(10,15\%,2,2\right)$ & 0.834 & 0.805 & 0.832 & 0.818 & \textbf{0.814} \\
			$\left(100,15\%,2,2\right)$ & 0.802 & 0.733 & 0.741 & 0.707 & \textbf{0.717} \\
			$\left(1,30\%,2,2\right)$ & 0.893 & 0.849 & 0.901 & 0.884 & \textbf{0.871} \\
			$\left(10,30\%,2,2\right)$ & 0.793 & 0.815 & 0.740 & 0.794 & \textbf{0.777} \\
			$\left(100,30\%,2,2\right)$ & 0.748 & 0.670 & 0.705 & 0.666 & \textbf{0.669} \\
			$\left(10,20\%,1.5,1\right)$ & 0.791 & 0.768 & 0.806 & 0.794 & \textbf{0.786} \\
			$\left(10,40\%,2.5,4\right)$ & 0.770 & 0.664 & 0.816 & 0.789 & \textbf{0.743} \\
			\bottomrule
		\end{tabular}
		\caption{$F_L$}
		\label{R2_tab_FL}
		\vspace{0.5cm}
	\end{subtable}
	\begin{subtable}{\textwidth}
		\centering
		\begin{tabular}{cccccc} 
			\toprule
			\multirow{2}{*}{$\left(\Re,\phi,d_l^*/d_s^*,V_l^*/V_s^*\right)$} & \multicolumn{2}{c}{Large spheres $R^2$} & \multicolumn{2}{c}{Small spheres $R^2$} & 	\multirow{2}{*}{Overall $R^2$} \\
			\cmidrule(lr){2-3}\cmidrule(lr){4-5}
			~ & Training sets & Test sets & Training sets & Test sets & ~ \\
			\midrule
			$\left(1,15\%,2,2\right)$ & 0.915 & 0.910 & 0.914 & 0.911 & \textbf{0.910} \\
			$\left(10,15\%,2,2\right)$ & 0.901 & 0.892 & 0.896 & 0.891 & \textbf{0.891} \\
			$\left(100,15\%,2,2\right)$ & 0.782 & 0.759 & 0.746 & 0.731 & \textbf{0.741} \\
			$\left(1,30\%,2,2\right)$ & 0.910 & 0.902 & 0.896 & 0.888 & \textbf{0.895} \\
			$\left(10,30\%,2,2\right)$ & 0.906 & 0.895 & 0.888 & 0.882 & \textbf{0.886} \\
			$\left(100,30\%,2,2\right)$ & 0.788 & 0.767 & 0.754 & 0.742 & \textbf{0.750} \\
			$\left(10,20\%,1.5,1\right)$ & 0.904 & 0.896 & 0.896 & 0.892 & \textbf{0.893} \\
			$\left(10,40\%,2.5,4\right)$ & 0.870 & 0.855 & 0.875 & 0.869 & \textbf{0.864} \\
			\bottomrule
		\end{tabular}
		\caption{$T_{\perp}$}
		\label{R2_tab_T}
	\end{subtable}			
	\caption{$R^2$ scores of $F_x$, $F_L$ and $T_{\perp}$ predictions for large and small spheres, as well as the overall performance of each dataset.}
	\label{R2_tab}
\end{table}

Finally, we analyze some statistical features of the $F_x$, $F_L$ and $T_{\perp}$ distributions obtained with PR-DNS and with the PINN model. We first compare the mean value of $F_x$ that is the major concern of classic drag models proposed in many previous works \cite{hoef_beetstra_kuipers_2005,Beetstra2007,Yin2009,Sarkar2009,Holloway2010,Cello2010,Rong2014,Mehrabadi2016,Duan2020}. It can be seen through the comparison that the largest error between the PR-DNS results and the PINN predictions is merely 0.18\%. Such a small discrepancy is undoubtedly better than that obtained with any existing drag closure. As mentioned previously, although we are not able to obtain a perfect prediction on each sphere, the errors are ideally centered around 0 and follow a Gaussian distribution, which compensates the positive errors with negative errors after summing over all spheres and ultimately yields extraordinary predictions of average $F_x$ that are close to the ground truth. Then we move our attention to the standard deviation of all three components. Overall, the standard deviations of all distributions are accurately predicted by our PINN model. The ratios of the largest to the smallest standard deviation of $F_x$, $F_L$ and $T_{\perp}$ are $7.4\sim7.5$, $6.3\sim6.7$ and $2.7\sim2.8$ respectively, confirming that $T_{\perp}$ is the least sensitive component to the variation of the key parameters. Last but not the least, we can observe that all standard deviations obtained with our PINN model are smaller than that computed by PR-DNS, demonstrating that the present PINN model is conservative in low data density regimes. This seems to imply that our model is unsurprisingly well trained for interpolation but not for extrapolation (please note that this is the common challenge for NN to deal with the unseen data).

\begin{table}
	\centering
	\renewcommand{\arraystretch}{1.2}
	\begin{tabular}{ccccccc} 
		\toprule
		\multirow{2}{*}{$\left(\Re,\phi,d_l^*/d_s^*,V_l^*/V_s^*\right)$} & \multicolumn{3}{c}{Large spheres} & \multicolumn{3}{c}{Small spheres} \\
		\cmidrule(lr){2-4}\cmidrule(lr){5-7}
		~ & $\langle F_{x,DNS}\rangle$ & $\langle F_{x,PINN}\rangle$ & Error & $\langle F_{x,DNS}\rangle$ & $\langle F_{x,PINN}\rangle$ & Error \\
		\midrule
		$\left(1,15\%,2,2\right)$ & 5.585 & 5.588 & 0.05\% & 2.918 & 2.914 & 0.12\% \\
		$\left(10,15\%,2,2\right)$ & 6.763 & 6.772 & 0.14\% & 3.509 & 3.512 & 0.10\% \\
		$\left(100,15\%,2,2\right)$ & 15.592 & 15.581 & 0.07\% & 7.673 & 7.665 & 0.11\% \\
		$\left(1,30\%,2,2\right)$ & 15.771 & 15.756 & 0.10\% & 6.712 & 6.711 & 0.00\% \\
		$\left(10,30\%,2,2\right)$ & 18.114 & 18.125 & 0.06\% & 7.701 & 7.701 & 0.01\% \\
		$\left(100,30\%,2,2\right)$ & 37.809 & 37.741 & 0.18\% & 15.784 & 15.795 & 0.07\% \\
		$\left(10,20\%,1.5,1\right)$ & 8.442 & 8.435 & 0.08\% & 5.491 & 5.483 & 0.14\% \\
		$\left(10,40\%,2.5,4\right)$ & 34.065 & 34.076 & 0.03\% & 10.085 & 10.081 & 0.04\% \\
		\bottomrule
	\end{tabular}
	\caption{Comparison of the mean values of $F_x$ for each case obtained with our PR-DNS and PINN predictions.}
	\label{mean_Fx_tab}
\end{table}

\begin{table}
	\centering
	\renewcommand{\arraystretch}{1.2}
	\begin{subtable}{\textwidth}
		\centering
		\begin{tabular}{ccccccc} 
			\toprule
			$\left(\Re,\phi,d_l^*/d_s^*,V_l^*/V_s^*\right)$ & $\sigma_{F_{x,DNS}}$ & $\sigma_{F_{x,PINN}}$ & $\sigma_{F_{L,DNS}}$ & $\sigma_{F_{L,PINN}}$ & $\sigma_{T_{\perp,DNS}}$ & $\sigma_{T_{\perp,PINN}}$ \\
			\midrule
			$\left(1,15\%,2,2\right)$ & 0.946 & 0.889 & 0.697 & 0.640 & 1.525 & 1.457 \\
			$\left(10,15\%,2,2\right)$ & 1.224 & 1.139 & 0.867 & 0.792 & 1.575 & 1.491 \\
			$\left(100,15\%,2,2\right)$ & 3.314 & 2.901 & 2.207 & 1.970 & 2.029 & 1.789 \\
			$\left(1,30\%,2,2\right)$ & 2.502 & 2.352 & 1.652 & 1.473 & 2.900 & 2.767 \\
			$\left(10,30\%,2,2\right)$ & 2.885 & 2.716 & 1.976 & 1.755 & 2.974 & 2.823 \\
			$\left(100,30\%,2,2\right)$ & 6.971 & 6.228 & 4.648 & 4.014 & 3.940 & 3.491 \\
			$\left(10,20\%,1.5,1\right)$ & 1.485 & 1.375 & 1.090 & 0.960 & 1.952 & 1.855 \\
			$\left(10,40\%,2.5,4\right)$ & 5.766 & 5.472 & 3.298 & 2.862 & 4.128 & 3.849 \\
			\bottomrule
		\end{tabular}
		\caption{Large spheres}
		\label{std_tab_large}
		\vspace{0.5cm}
	\end{subtable}
	\begin{subtable}{\textwidth}
		\centering
		\begin{tabular}{ccccccc} 
			\toprule
			$\left(\Re,\phi,d_l^*/d_s^*,V_l^*/V_s^*\right)$ & $\sigma_{F_{x,DNS}}$ & $\sigma_{F_{x,PINN}}$ & $\sigma_{F_{L,DNS}}$ & $\sigma_{F_{L,PINN}}$ & $\sigma_{T_{\perp,DNS}}$ & $\sigma_{T_{\perp,PINN}}$ \\
			\midrule
			$\left(1,15\%,2,2\right)$ & 0.681 & 0.650 & 0.529 & 0.497 & 1.126 & 1.077 \\
			$\left(10,15\%,2,2\right)$ & 0.950 & 0.897 & 0.641 & 0.584 & 1.169 & 1.106 \\
			$\left(100,15\%,2,2\right)$ & 2.545 & 2.195 & 1.457 & 1.255 & 1.404 & 1.208 \\
			$\left(1,30\%,2,2\right)$ & 1.626 & 1.543 & 1.223 & 1.126 & 2.247 & 2.128 \\
			$\left(10,30\%,2,2\right)$ & 2.013 & 1.906 & 1.466 & 1.320 & 2.346 & 2.202 \\
			$\left(100,30\%,2,2\right)$ & 5.118 & 4.605 & 3.314 & 2.784 & 2.954 & 2.565 \\
			$\left(10,20\%,1.5,1\right)$ & 1.308 & 1.221 & 0.929 & 0.833 & 1.659 & 1.571 \\
			$\left(10,40\%,2.5,4\right)$ & 3.000 & 2.841 & 2.170 & 1.956 & 3.122 & 2.920 \\
			\bottomrule
		\end{tabular}
		\caption{Small spheres}
		\label{std_tab_small}
	\end{subtable}
	\caption{Comparison of the standard deviations of $F_x$, $F_L$ and $T_{\perp}$ for each case obtained with our PR-DNS and PINN predictions.}
	\label{std_tab}
\end{table}

\subsection{Extension to a universal model}
	
	In the previous subsection, we trained and evaluated our PINN model on each individual dataset as summarized in Table \ref{R2_tab}. It is natural to wonder if a single model could provide predictions at the same level of accuracy in a broad parameter space and would hence not be limited to a specific configuration, as the ultimate objective of this study is to find a promising alternative to the average drag closures that are typically applicable over a range of $\Re$, $\phi$,  $d_l^*/d_s^*$ and $V_l^*/V_s^*$. Therefore, we implement the training procedures for the $F_x$ prediction using the same hyper-parameters as those optimized in Section \ref{subsec:hyper-pm} on multiple integrated datasets, and run additional simulations with distinct sets of parameters that fall within the input space described in Table \ref{prams_tab}. We feed the new datasets into the updated and so called universal PINN model and evaluate the corresponding $R^2$ scores, thus examining the interpolation capability of the present model within a given parameter regime. 
	
	\begin{figure}
		\centering
		\captionsetup[subfigure]{oneside,margin={0.5cm,0.cm}}
		\begin{subfigure}[c]{0.90\linewidth}		
			\resizebox{\linewidth}{!}{\includegraphics{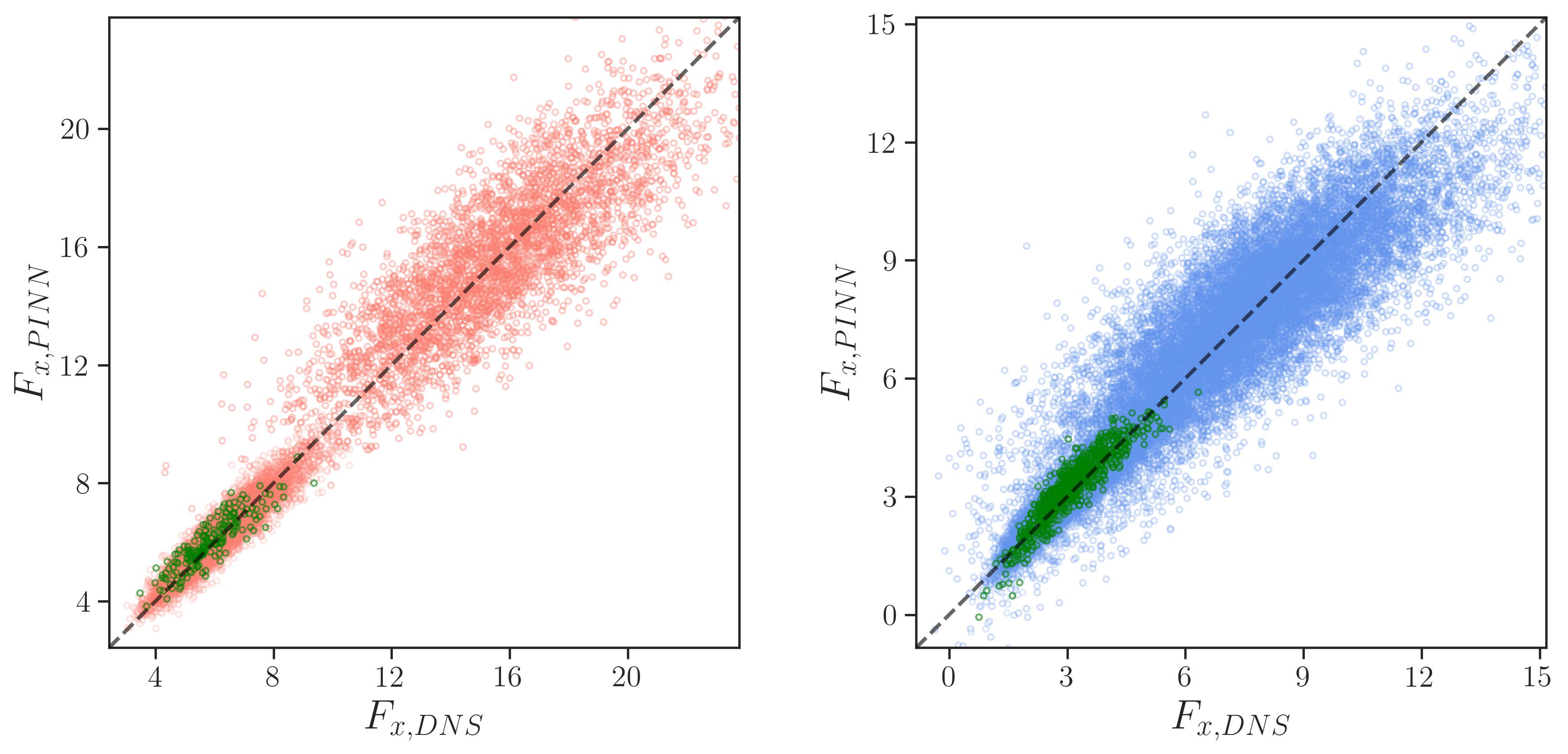}}
			\vspace{-0.7cm}
			\subcaption{Interpolation of $\Re$}
			\vspace{0.5cm}
			\label{Uni_Re}
		\end{subfigure}
		\begin{subfigure}[c]{0.90\linewidth}		
			\resizebox{\linewidth}{!}{\includegraphics{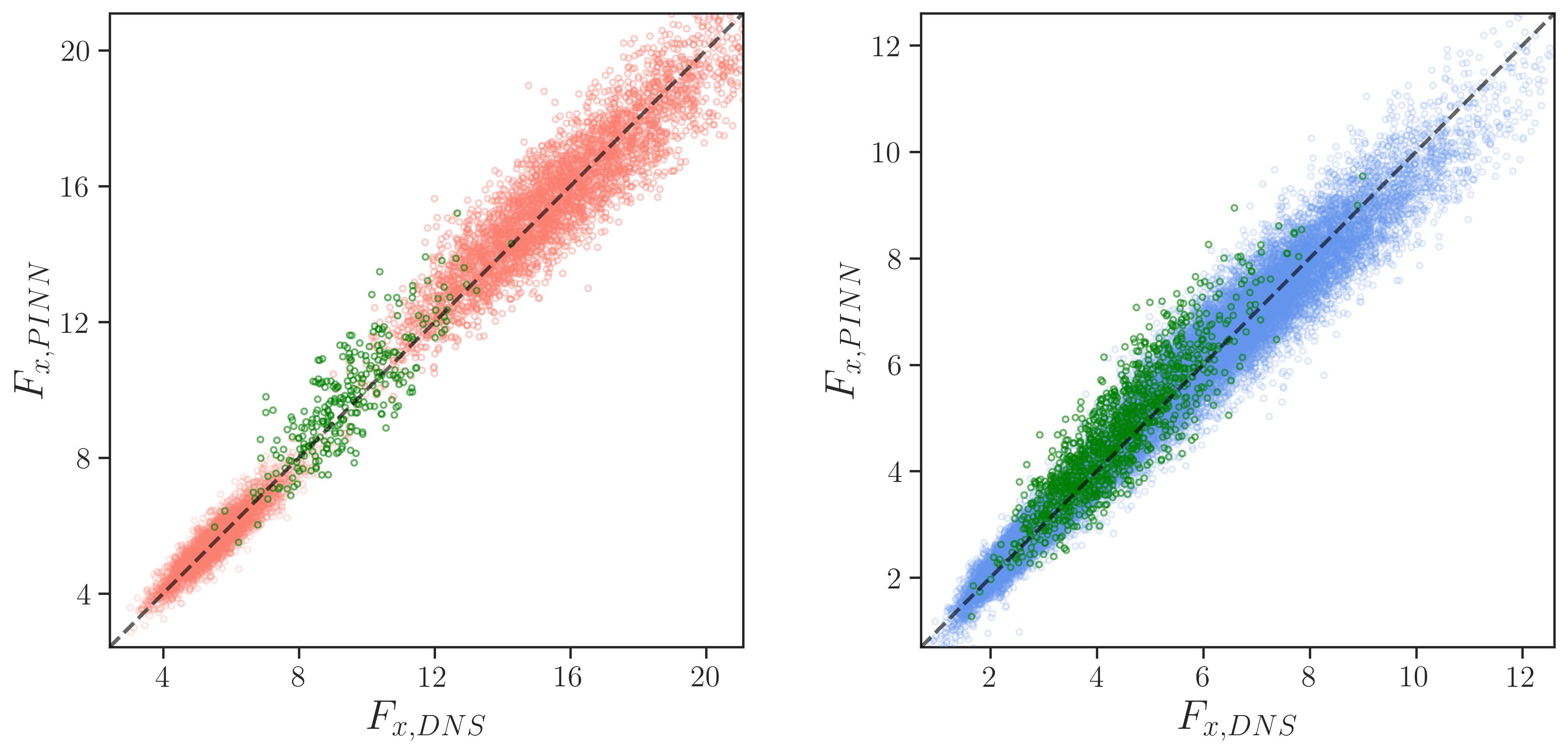}}
			\vspace{-0.7cm}
			\subcaption{Interpolation of $\phi$}
			\vspace{0.5cm}
			\label{Uni_phi}
		\end{subfigure}
		\begin{subfigure}[c]{0.90\linewidth}		
			\resizebox{\linewidth}{!}{\includegraphics{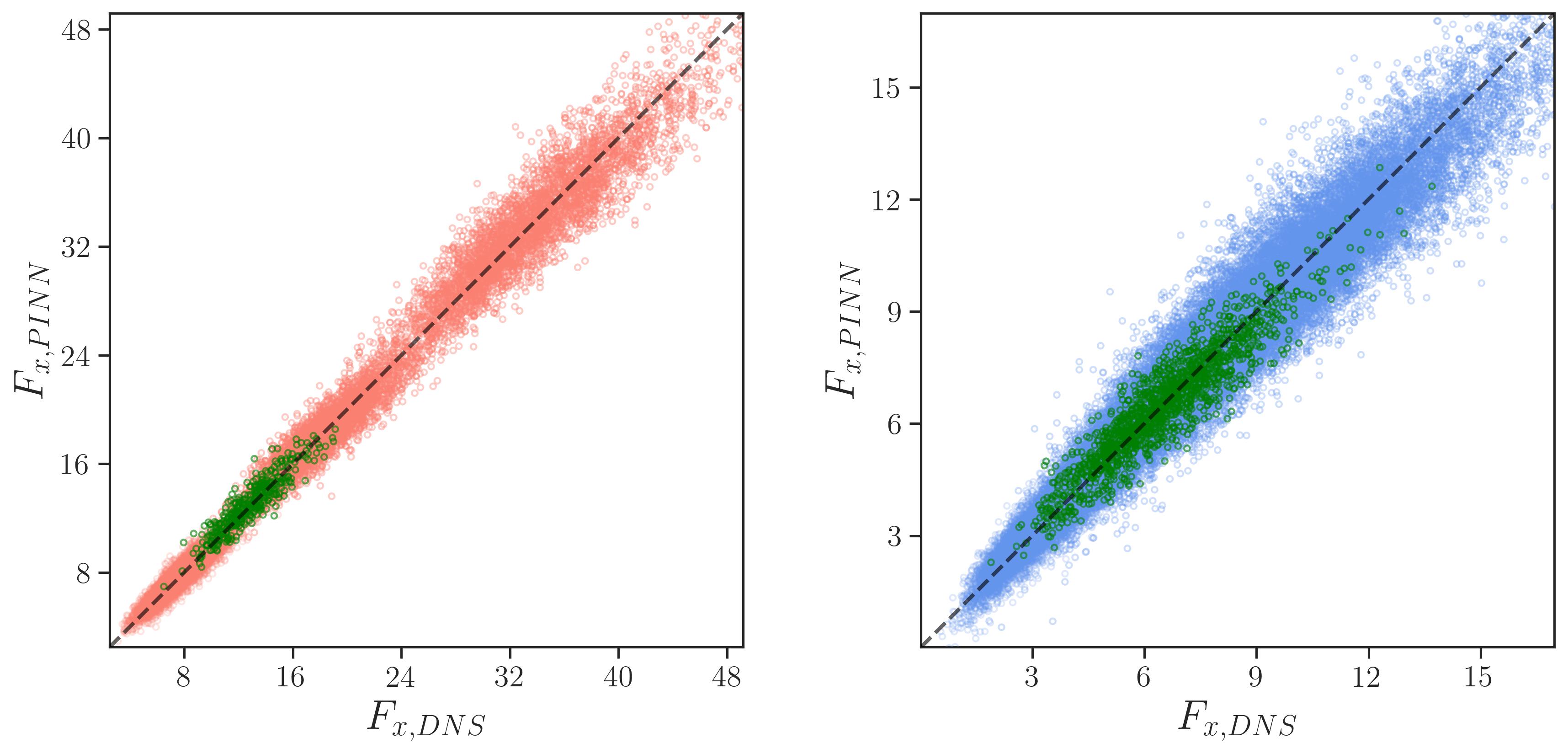}}
			\vspace{-0.7cm}
			\subcaption{Interpolation of all particle properties}
			\label{Uni_geo}
		\end{subfigure}
		\caption{Prediction on the unseen data (green dots) using the PINN models trained on integrated datasets (red and blue dots for large and small spheres respectively).}
		\label{Uni_inter_plot}
	\end{figure}
	
	\begin{table}
		\centering
		\renewcommand{\arraystretch}{1.2}
		\begin{subtable}{\textwidth}
			\centering
			\begin{tabular}{ccccc} 
				\toprule
				\multirow{2}{*}{$\left(\Re,\phi,d_l^*/d_s^*,V_l^*/V_s^*\right)$} & \multicolumn{2}{c}{Large spheres $R^2$} & \multicolumn{2}{c}{Small spheres $R^2$} \\
				\cmidrule(lr){2-3}\cmidrule(lr){4-5}
				~ & Training sets & Test sets & Training sets & Test sets\\
				\midrule
				$\left(1,15\%,2,2\right)$ & 0.878 & 0.862 & 0.895 & 0.890\\
				$\left(10,15\%,2,2\right)$ & 0.859 & 0.842 & 0.862 & 0.865\\
				$\left(100,15\%,2,2\right)$ & 0.754 & 0.719 & 0.721 & 0.704\\	\cdashline{1-5}	
				$\left(5,15\%,2,2\right)$ & - & 0.764 & - & 0.805 \\
				\bottomrule
			\end{tabular}
			\caption{Interpolation of $\Re$}
			\label{Uni_inter_Re}
			\vspace{0.5cm}
		\end{subtable}
		\begin{subtable}{\textwidth}
			\centering
			\begin{tabular}{cccccc} 
				\toprule
				\multirow{2}{*}{$\left(\Re,\phi,d_l^*/d_s^*,V_l^*/V_s^*\right)$} & \multicolumn{2}{c}{Large spheres $R^2$} & \multicolumn{2}{c}{Small spheres $R^2$} \\
				\cmidrule(lr){2-3}\cmidrule(lr){4-5}
				~ & Training sets & Test sets & Training sets & Test sets\\
				\midrule
				$\left(1,15\%,2,2\right)$ & 0.882 & 0.869 & 0.895 & 0.895\\
				$\left(1,30\%,2,2\right)$ & 0.880 & 0.853 & 0.895 & 0.887\\ \cdashline{1-5}
				$\left(1,22.5\%,2,2\right)$ & - & 0.535 & - & 0.603 \\
				\bottomrule
			\end{tabular}
			\caption{Interpolation of $\phi$}
			\label{Uni_inter_phi}
			\vspace{0.5cm}
		\end{subtable}
		\begin{subtable}{\textwidth}
			\centering
			\begin{tabular}{cccccc} 
				\toprule
				\multirow{2}{*}{$\left(\Re,\phi,d_l^*/d_s^*,V_l^*/V_s^*\right)$} & \multicolumn{2}{c}{Large spheres $R^2$} & \multicolumn{2}{c}{Small spheres $R^2$} \\
				\cmidrule(lr){2-3}\cmidrule(lr){4-5}
				~ & Training sets & Test sets & Training sets & Test sets\\
				\midrule
				$\left(10,15\%,2,2\right)$ & 0.852 & 0.846 & 0.862 & 0.863\\
				$\left(10,20\%,1.5,1\right)$ & 0.835 & 0.831 & 0.863 & 0.862\\
				$\left(10,30\%,2,2\right)$ & 0.857 & 0.848 & 0.881 & 0.878\\
				$\left(10,40\%,2.5,4\right)$ & 0.893 & 0.867 & 0.886 & 0.873\\
				\cdashline{1-5}
				$\left(10,25\%,1.8,1.5\right)$ & - & 0.833 & - & 0.862\\
				\bottomrule
			\end{tabular}
			\caption{Interpolation of all particle properties}
			\label{Uni_inter_geo}
		\end{subtable}			
		\caption{$R^2$ scores of $F_x$ predictions for large and small spheres on both training and unseen datasets.}
		\label{Uni_inter_tab}
	\end{table} 
	
	There are four changing parameters in our input space: flow property $\Re$ and particle properties $\phi$, $d_l^*/d_s^*$ and $V_l^*/V_s^*$. We start with the simplest scenario of interpolating in a single direction while fixing all the other parameters, and subsequently proceed to a more challenging case of interpolating all particle properties:
	\begin{enumerate}
		\item Interpolation of $\Re$: we train the PINN model on three datasets with the same particle properties  $\left(\phi,d_l^*/d_s^*,V_l^*/V_s^*\right)=\left(15\%,2,2\right)$ at $\Re=1$, 10 and 100, and generate an additional dataset with the same particle properties at $\Re=5$. We exhibit the correlation plot of all PR-DNS data and PINN predictions in Figure \ref{Uni_inter_plot} where the large and small particles with seen parameters (i.e., $\Re=1$, 10 and 100 that are used for training the model) are displayed in red and blue respectively, and predictions with unseen parameter (i.e., $\Re=5$ that does not contribute to the training process) are represented by green dots. We can see from Figure \ref{Uni_Re} that these green dots are distributed around the black diagonal line that corresponds to $R^2=1$, indicating good performance in predicting unseen data at $\Re=5$. The quantitative results are presented in Table \ref{Uni_inter_tab} where similar $R^2$ scores can be observed for training datasets as those reported in Table \ref{R2_tab} produced by the separate models, and good predictions on the unseen dataset can be also claimed with $R^2=0.764$ and $0.805$ for large and small spheres respectively.
		\item Interpolation of $\phi$: we train the PINN model on two datasets with the same $\left(\Re,d_l^*/d_s^*,V_l^*/V_s^*\right)=\left(1,2,2\right)$ but different $\phi=15\%$ and 30\%, and generate an extra dataset at $\phi=22.5\%$. Figure \ref{Uni_phi} reveals that green dots (predictions on unseen data) are loosely distributed about the diagonal line, implying a higher level of prediction error in general, with overestimation occurring most frequently for those representing larger $F_x$. Quantitatively, the $R^2$ decreases to $0.535$ for large spheres and $0.603$ for small spheres when only two datasets (lower and upper bounds with a broad interval) are considered when training the model, which is a compromised choice due to the limited number of data available in the present study. 
		\item Interpolation of all particle properties: we train the PINN model on four datasets with the same $\Re=10$ but various particle properties, and then generate a new dataset with $\left(\phi,d_l^*/d_s^*,V_l^*/V_s^*\right)=\left(25\%,1.8,1.5\right)$ such that all three parameters lie within the input ranges and differ to any original inputs (please note that we add $V_l^*/V_s^*$ to the input space to train this model). Figure \ref{Uni_geo} demonstrates astonishingly accurate predictions on the unseen data and the corresponding $R^2=0.833$ and $0.862$ can be found in Table \ref{Uni_inter_geo} for large and small spheres respectively, which are even very close to the $R^2$ scores of the training datasets. One plausible explanation is that the increments in the input parameters of these four datasets are smaller than those of previous two tests ($\Re=1\rightarrow10\rightarrow100$ and $\phi=15\%\rightarrow30\%$), resulting in a more uniform distribution of training data that leaves fewer and smaller unexplored regions in the parameter space as shown in Figure \ref{Uni_geo}. 
	\end{enumerate}   
	
	\begin{figure}
		\centering
		\captionsetup[subfigure]{oneside,margin={0.5cm,0.cm}}
		\begin{subfigure}[c]{0.90\linewidth}		
			\resizebox{\linewidth}{!}{\includegraphics{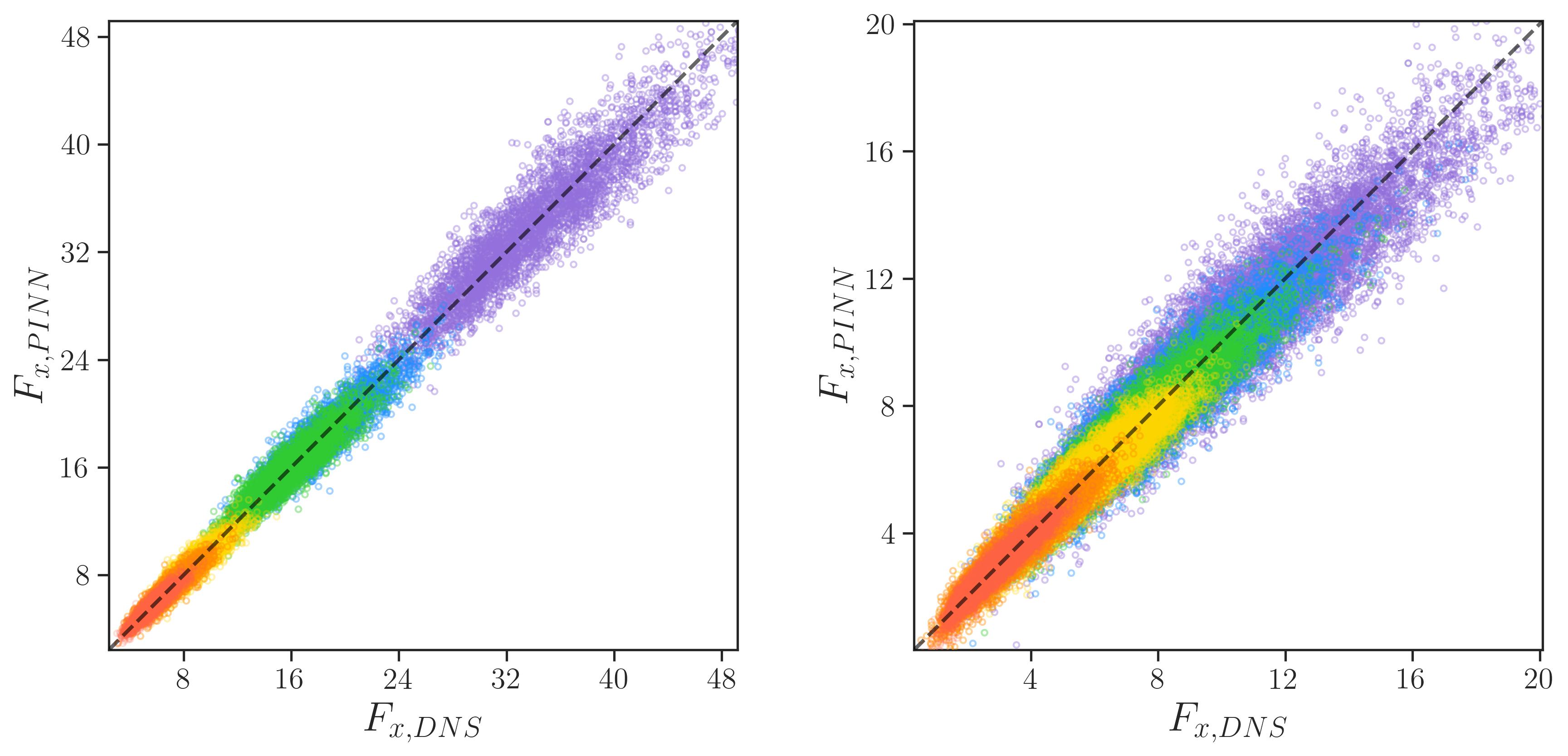}}
			\vspace{-0.7cm}
			\subcaption{Universal model}
			\vspace{0.5cm}
			\label{Uni_model}
		\end{subfigure}
		\begin{subfigure}[c]{0.90\linewidth}		
			\resizebox{\linewidth}{!}{\includegraphics{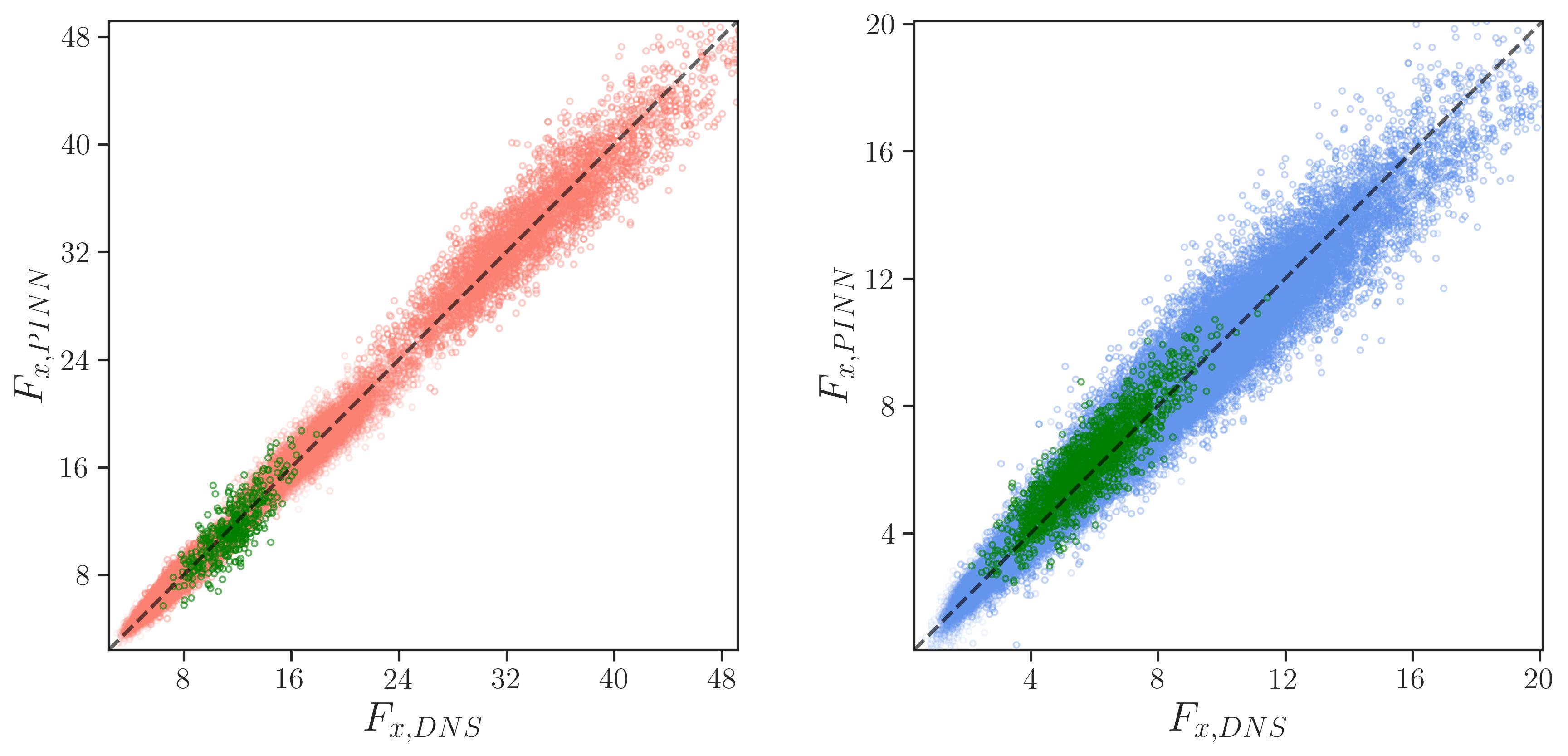}}
			\vspace{-0.7cm}
			\subcaption{Predictions on unseen data}
			\label{Uni_inter}
		\end{subfigure}
		\caption{Training of a universal model. (a) Comparison of $F_x$ obtained with our PR-DNS and universal PINN predictions. Left: large spheres, and right: small spheres. (b) Interpolation of all parameters.}
		\label{Uni_plot}
	\end{figure}
	
	\begin{table}
		\centering
		\renewcommand{\arraystretch}{1.2}
		\centering
		\begin{tabular}{ccccc} 
			\toprule
			\multirow{2}{*}{$\left(\Re,\phi,d_l^*/d_s^*,V_l^*/V_s^*\right)$} & \multicolumn{2}{c}{Large spheres $R^2$} & \multicolumn{2}{c}{Small spheres $R^2$} \\
			\cmidrule(lr){2-3}\cmidrule(lr){4-5}
			~ & Training sets & Test sets & Training sets & Test sets\\
			\midrule
			$\cellcolor{red!40}\left(1,15\%,2,2\right)$ & 0.868 & 0.859 & 0.885 & 0.880\\
			$\cellcolor{orange!50}\left(10,15\%,2,2\right)$ & 0.861 & 0.851 & 0.865 & 0.867\\
			$\cellcolor{yellow!40}\left(10,20\%,1.5,1\right)$ & 0.838 & 0.832 & 0.863 & 0.862\\
			$\cellcolor{green!40}\left(1,30\%,2,2\right)$ & 0.870 & 0.858 & 0.891 & 0.887\\
			$\cellcolor{cyan!50}\left(10,30\%,2,2\right)$ & 0.863 & 0.850 & 0.882 & 0.879\\
			$\cellcolor{blue!30}\left(10,40\%,2.5,4\right)$ & 0.894 & 0.874 & 0.887 & 0.879\\
			\cdashline{1-5}
			$\left(5,25\%,1.8,1.5\right)$ & - & 0.459 & - & 0.714\\
			\bottomrule
		\end{tabular}
		\caption{$R^2$ scores of large and small spheres in various datasets with a universal PINN model.}
		\label{Uni_tab}
	\end{table}
	
	Given the excellent performance of our PINN modelling on multiple datasets and its demonstrated interpolation capability to the unseen data, we are now in the position to establish a universal model that incorporates both variable fluid and particle properties. With the same logic of avoiding large increments in parameter space, we omit the datasets with $\Re=100$ and combine the rest with $\Re=1$ and $\Re=10$ comprising over $110,000$ data points. We plot all the PR-DNS data with their corresponding prediction values obtained by our universal PINN model in Figure \ref{Uni_model}, and list the $R^2$ scores for all the six training datasets in Table \ref{Uni_tab} (please find each case and its corresponding clustered scatters according to the colors illustrated in the figure and table). Throughout the comparison of $F_x$ predictions generated by the separate and universal models as shown in Table \ref{R2_tab} and \ref{Uni_tab} respectively, we can observe that the universal model accomplishes great performance which is highly similar to the model constructed with independent training. For the cases with relatively larger mean values such as $\left(\Re,\phi,d_l^*/d_s^*,V_l^*/V_s^*\right)=\left(10,30\%,2,2\right)$ and $\left(10,40\%,2.5,4\right)$, our universal model produces even slightly better predictions for both the large and small spheres, which is not surprising given that (\expandafter{\romannumeral1}) more high-quality data are provided into the PINN model, and (\expandafter{\romannumeral2}) $\delta$ for evaluating the Huber loss is set as the largest standard deviation among these six datasets, i.e., that of large spheres in case of $\left(10,40\%,2.5,4\right)$. The success of such a universal model justifies our previous implementation of training and testing on each individual dataset, which significantly accelerates the fine-tuning process of hyper-parameters and offers a fast way to determine the model performance by merely examining the lower and upper bounds of parameter space. 
	
	We eventually generate a dataset with $\left(\Re,\phi,d_l^*/d_s^*,V_l^*/V_s^*\right)=\left(5,25\%,1.8,1.5\right)$ to examine the interpolation capability of our universal model in the entire input space. Figure \ref{Uni_inter} and Table \ref{Uni_tab} show the distribution and $R^2$ scores of predictions on this group of unseen data respectively, and it can be noted that the predictive performance, especially in the low $\Re$ regime, is acceptable qualitatively and quantitatively considering the complexity of this test, while we can reasonably expect much more accurate predictions that may be provided by a universal model trained on uniformly distributed data with higher density. Preliminary findings reported in the above reveal that interpolating in the $\phi$ dimension of the parameter space constitutes a major challenge and therefore high density of the training dataset in the $\phi$ direction should be favored. This is not a surprising recommendation from a Fluid Mechanics viewpoint since the level of force/torque fluctuations depend on the strength of the flow disturbances caused by neighbors, and the magnitude of these flow disturbances primarily depends on $\phi$.

\subsection{Interpretability of the PINN model}

\begin{figure}
	\centering
	\captionsetup[subfigure]{oneside,margin={0.cm,0.cm}}
	\begin{subfigure}[c]{0.8\linewidth}		
		\resizebox{\linewidth}{!}{\includegraphics{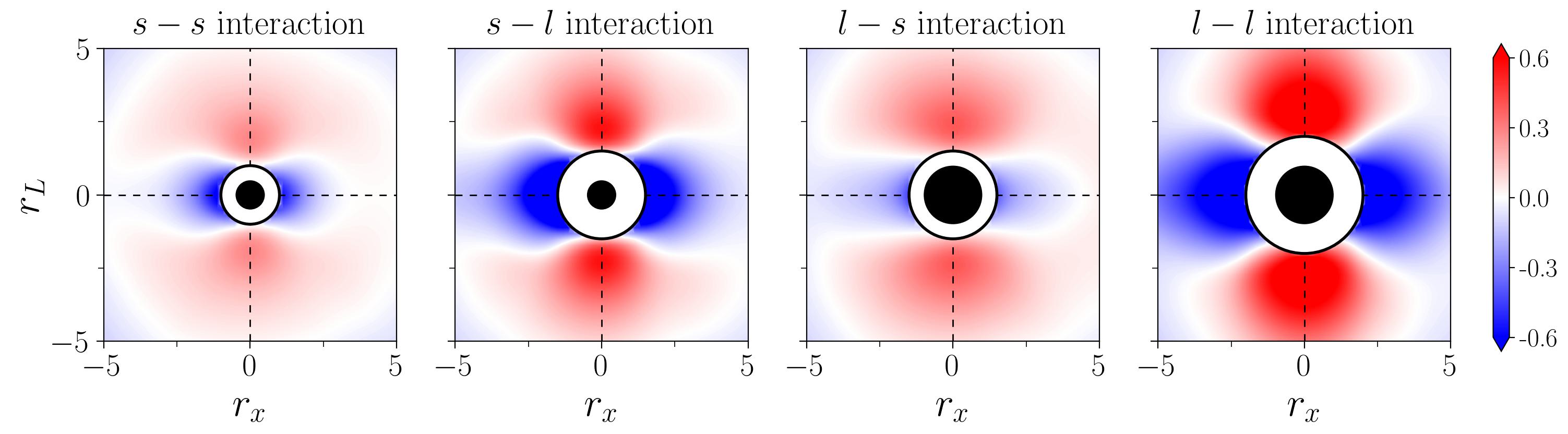}}
		\vspace{-0.7cm}
		\subcaption{$\left(Re,\phi,d_l^*/d_s^*,V_l^*/V_s^*\right)=(1,15\%,2,2)$}
		\vspace{0.cm}
		\label{bi_inter_Fx_Re1}
	\end{subfigure}
	\begin{subfigure}[c]{0.8\linewidth}		
		\resizebox{\linewidth}{!}{\includegraphics{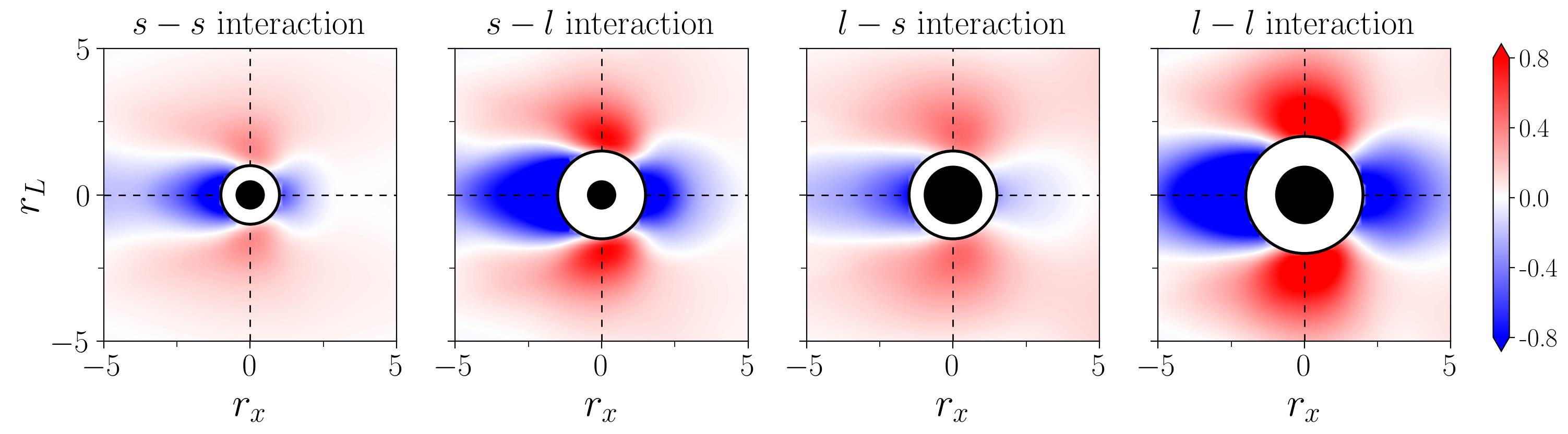}}
		\vspace{-0.7cm}
		\subcaption{$\left(Re,\phi,d_l^*/d_s^*,V_l^*/V_s^*\right)=(10,15\%,2,2)$}
		\label{bi_inter_Fx_Re10}
	\end{subfigure}
	\begin{subfigure}[c]{0.8\linewidth}		
		\resizebox{\linewidth}{!}{\includegraphics{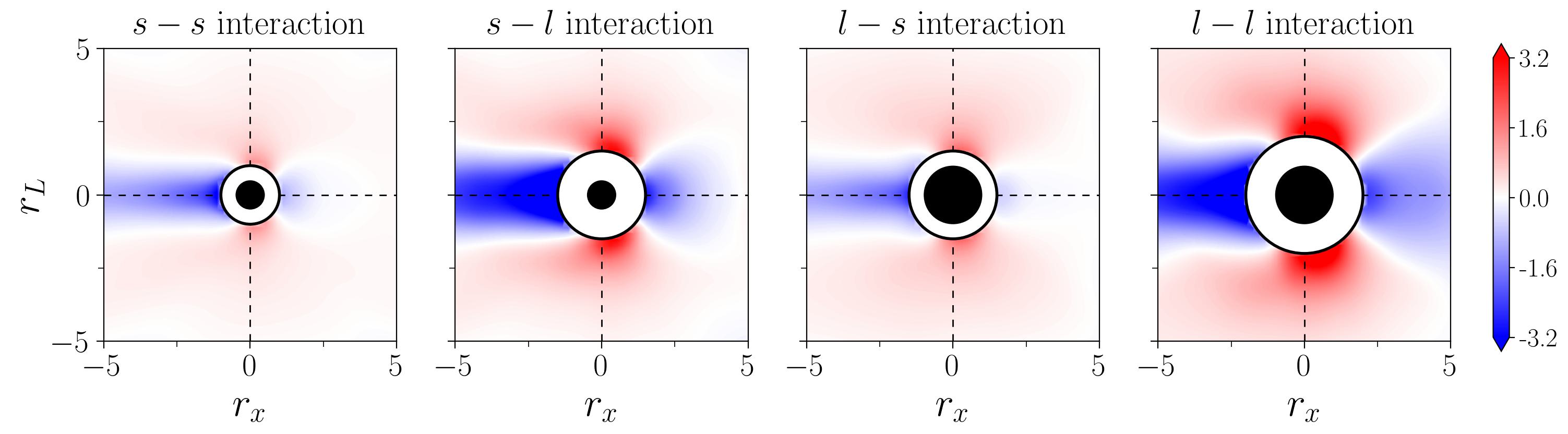}}
		\vspace{-0.7cm}
		\subcaption{$\left(Re,\phi,d_l^*/d_s^*,V_l^*/V_s^*\right)=(100,15\%,2,2)$}
		\label{bi_inter_Tx_Re100}
	\end{subfigure}
	\caption{$F_x$ contribution of the closest neighbor to the reference particle in different interaction modes.}
	\label{bi_inter_Fx}
\end{figure}

\begin{figure}
	\centering
	\captionsetup[subfigure]{oneside,margin={0.cm,0.cm}}
	\begin{subfigure}[c]{0.8\linewidth}		
		\resizebox{\linewidth}{!}{\includegraphics{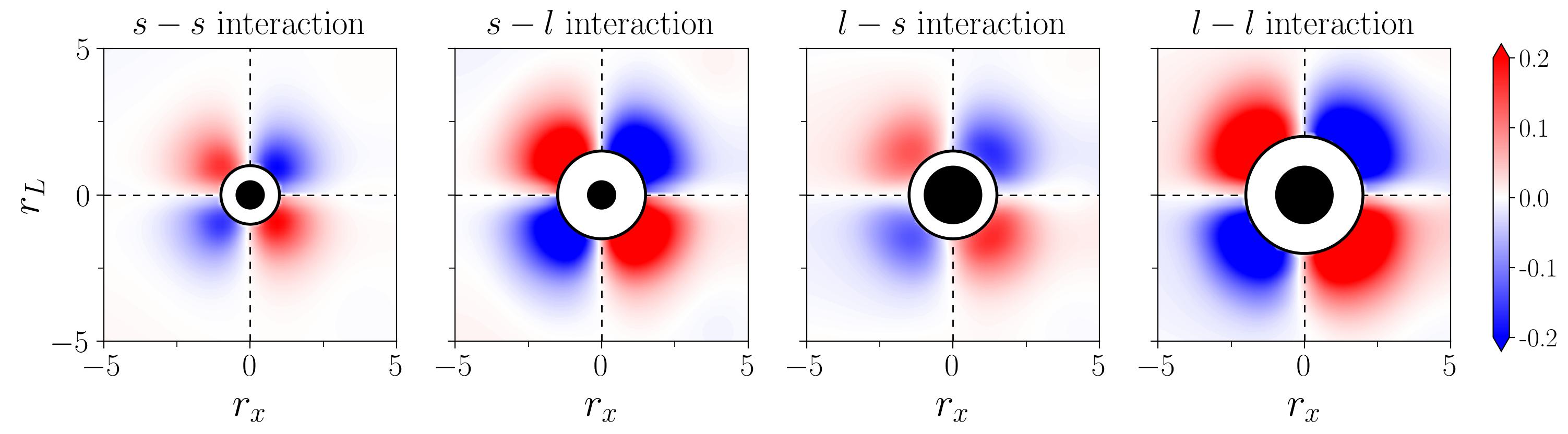}}
		\vspace{-0.7cm}
		\subcaption{$\left(Re,\phi,d_l^*/d_s^*,V_l^*/V_s^*\right)=(1,15\%,2,2)$}
		\vspace{0.cm}
		\label{bi_inter_FL_Re1}
	\end{subfigure}
	\begin{subfigure}[c]{0.8\linewidth}		
		\resizebox{\linewidth}{!}{\includegraphics{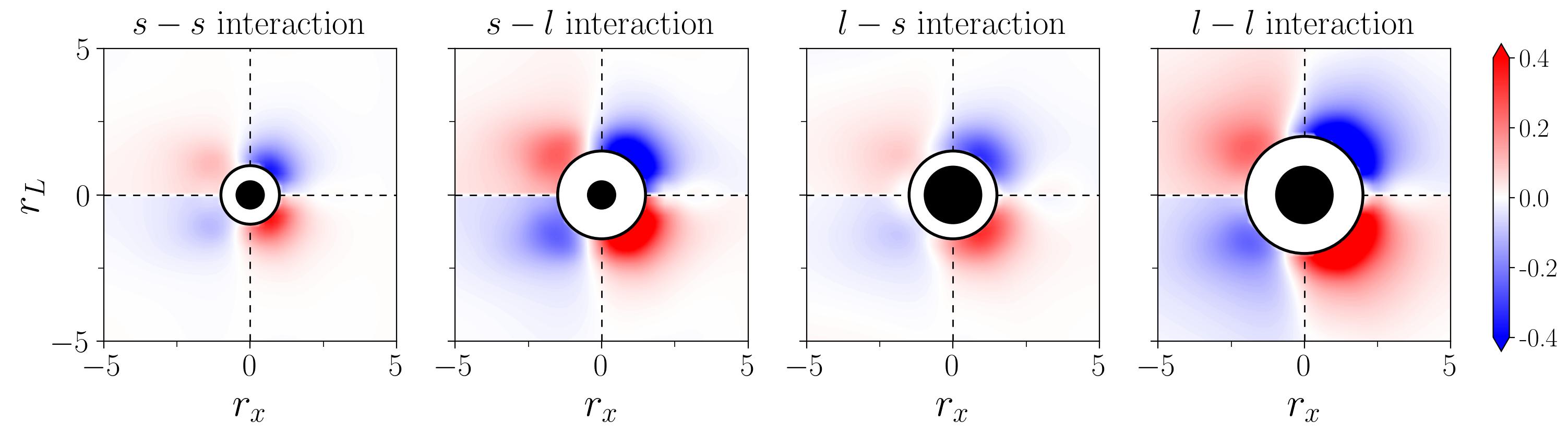}}
		\vspace{-0.7cm}
		\subcaption{$\left(Re,\phi,d_l^*/d_s^*,V_l^*/V_s^*\right)=(10,15\%,2,2)$}
		\label{bi_inter_FL_Re10}
	\end{subfigure}
	\begin{subfigure}[c]{0.8\linewidth}		
		\resizebox{\linewidth}{!}{\includegraphics{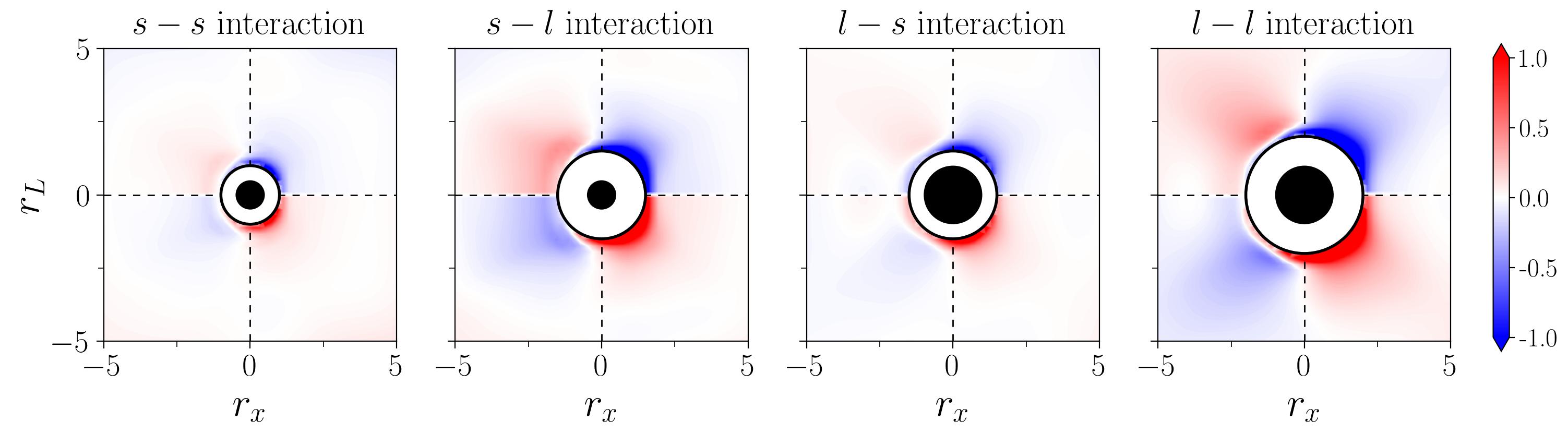}}
		\vspace{-0.7cm}
		\subcaption{$\left(Re,\phi,d_l^*/d_s^*,V_l^*/V_s^*\right)=(100,15\%,2,2)$}
		\label{bi_inter_FL_Re100}
	\end{subfigure}
	\caption{$F_L$ contribution of the closest neighbor to the reference particle in different interaction modes.}
	\label{bi_inter_FL}
\end{figure}

\begin{figure}
	\centering
	\captionsetup[subfigure]{oneside,margin={0.cm,0.cm}}
	\begin{subfigure}[c]{0.8\linewidth}		
		\resizebox{\linewidth}{!}{\includegraphics{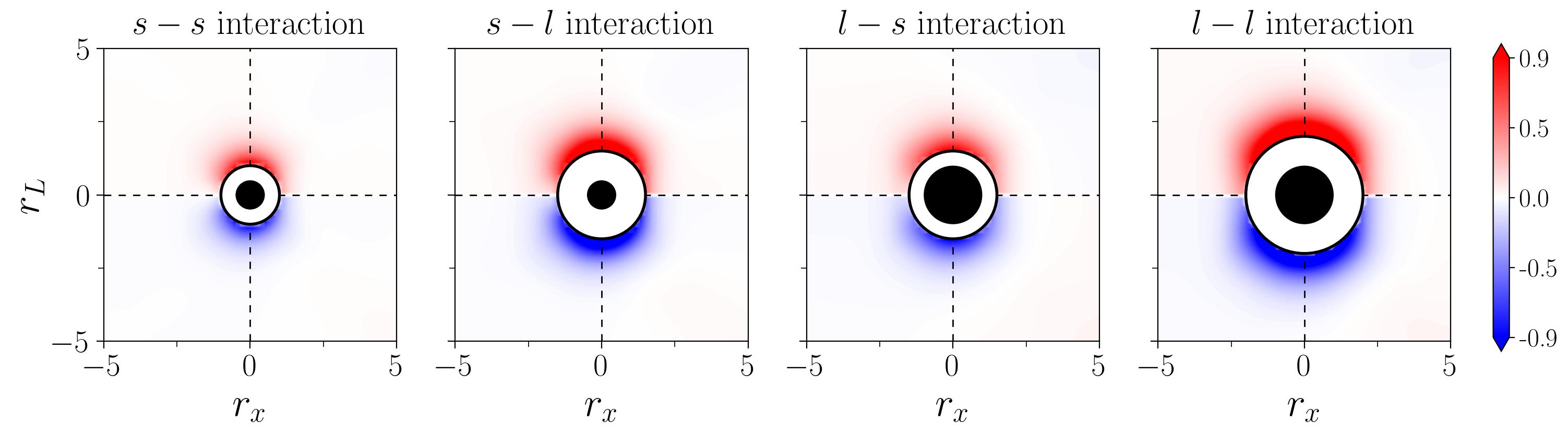}}
		\vspace{-0.7cm}
		\subcaption{$\left(Re,\phi,d_l^*/d_s^*,V_l^*/V_s^*\right)=(1,15\%,2,2)$}
		\vspace{0.cm}
		\label{bi_inter_TL_Re1}
	\end{subfigure}
	\begin{subfigure}[c]{0.8\linewidth}		
		\resizebox{\linewidth}{!}{\includegraphics{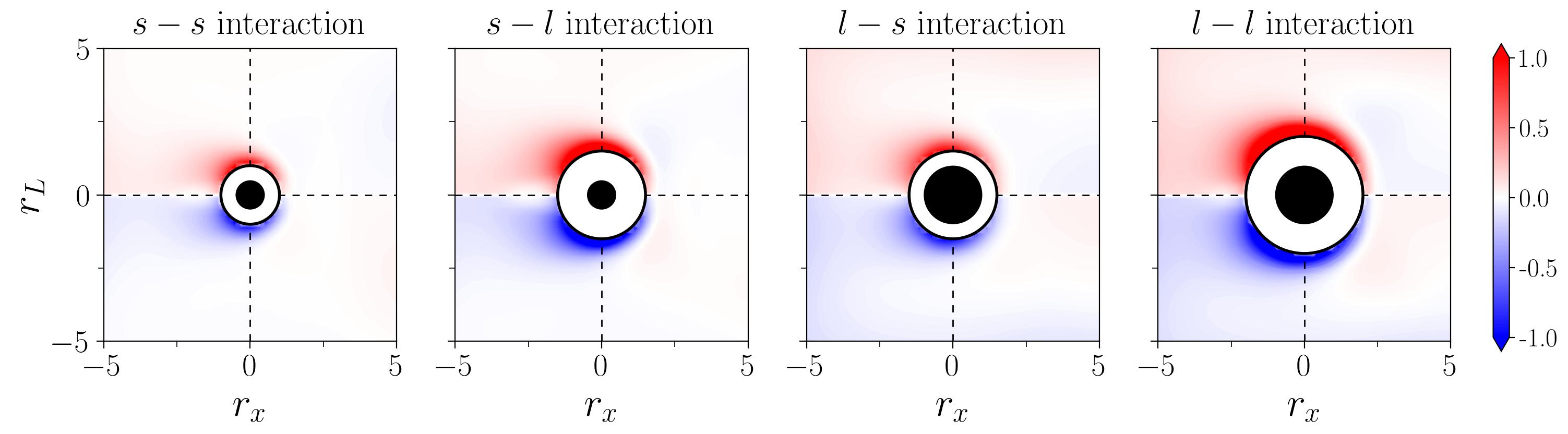}}
		\vspace{-0.7cm}
		\subcaption{$\left(Re,\phi,d_l^*/d_s^*,V_l^*/V_s^*\right)=(10,15\%,2,2)$}
		\label{bi_inter_TL_Re10}
	\end{subfigure}
	\begin{subfigure}[c]{0.8\linewidth}		
		\resizebox{\linewidth}{!}{\includegraphics{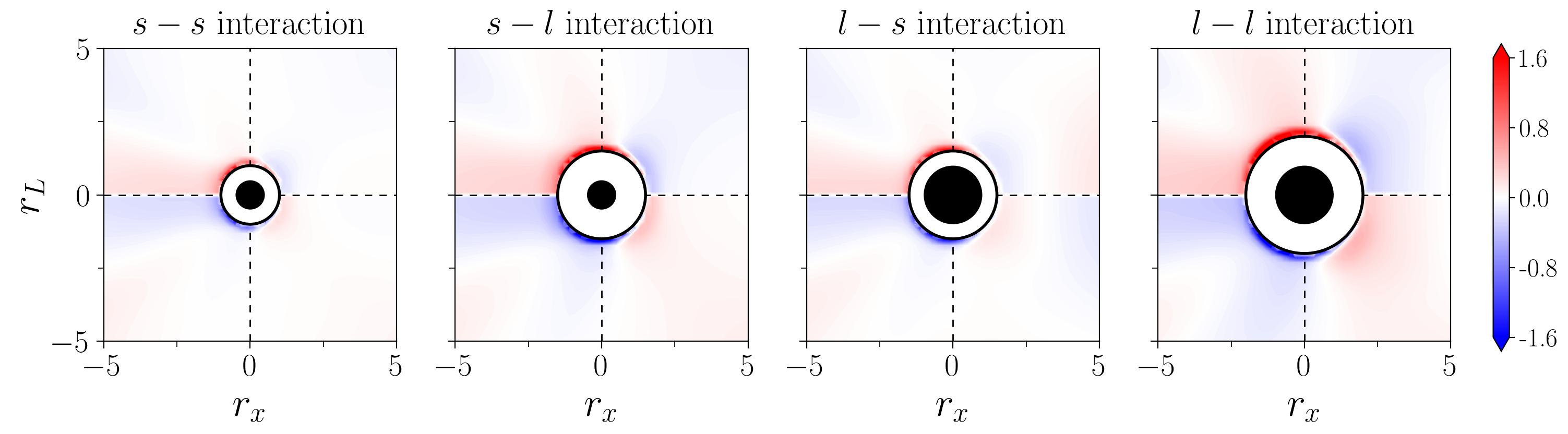}}
		\vspace{-0.7cm}
		\subcaption{$\left(Re,\phi,d_l^*/d_s^*,V_l^*/V_s^*\right)=(100,15\%,2,2)$}
		\label{bi_inter_TL_Re100}
	\end{subfigure}
	\caption{$T_{\perp}$ contribution of the closest neighbor to the reference particle in different interaction modes.}
	\label{bi_inter_TL}
\end{figure}

Apart from the generalization, interpretability is also a critical consideration in a NN model. It is of great importance for the model to learn the underlying mechanics and reproduce results that are physically meaningful. We borrow the idea from Seyed-Ahmadi and Wachs \cite{SeyedAhmadi2022} and generate the contribution from the nearest and second nearest neighbors to the total force/torque exerted on a target sphere. In fact, we can easily access the individual contribution of each neighbor giving to that our NN architecture separately evaluates every pairwise interaction and only sums over all influential neighbors at the output layer with a set of weights. We first present the contribution of $F_x$, $F_L$ and $T_{\perp}$ from the nearest neighbor in Figures \ref{bi_inter_Fx}$-$\ref{bi_inter_TL}. Please note that in these plots, we put the target sphere (solid black circle) at the center and the outer white space defines the region where the neighbor cannot reside. We identify four different interaction modes: $s-s$, $s-l$, $l-s$ and $l-l$ interactions in bidisperse particle-laden flows with the first letter representing the size of the target sphere and the second letter that of its neighbor ($l$: large or $s$: small). We adopt the same notations as in \cite{SeyedAhmadi2022} where $r_x$ stands for the streamwise distance between two spheres and $r_L$ for the transverse distance. The plots generated as follows are processed with an azimuthal average operation over several planes containing the streamwise $x$ direction (which is also the reason why we use $r_L$ instead of $r_y$ or $r_z$). The perfect symmetry about the $x$-axis in each plot is an outcome of data augmentation (we mirror the dataset about $x$-axis on the $xy$ and $xz$ planes due to the flow symmetry) but not of this average operation. For the illustration purpose, we select three cases  $\left(\Re,\phi,d_l^*/d_s^*,V_l^*/V_s^*\right)=\left(1,15\%,2,2\right)$, $\left(10,15\%,2,2\right)$ and $\left(100,15\%,2,2\right)$ while similar patterns can be observed in all other cases.

These contribution plots are not only similar to those presented in \cite{Moore2019,SeyedAhmadi2022,Siddani2023}, but also reminiscent of the probability distribution maps in \cite{SeyedAhmadi2020,Cheng2023}. As an example, the $F_x$ contributions in case $\left(100,15\%,2,2\right)$ illustrated in Figure \ref{bi_inter_Fx} are highly comparable to the combination of $\tilde{p}\left(\Delta F_x>0\right)$ and $\tilde{p}\left(\Delta F_x<0\right)$ in \cite{Cheng2023}, where the high probability regions of a target sphere experiencing lower than the average streamwise force are constantly located in mostly upstream and somewhat downstream areas, meanwhile it is shown from the contribution plots that a neighboring sphere induces a negative $\Delta F_x$ in the qualitatively identical regions (in blue). Similarly, the positive $\Delta F_x$ contribution areas (in red), i.e., the high probability regions of $\Delta F_x>0$ since the nearest neighbor plays a dominant role, are constantly located on both sides of a target sphere. It should be noted that the red regions are slightly drifted downstream-wise at both $\Re=10$ and $100$, while a fore-aft symmetry is captured at $\Re=1$ where the flow inertia has a minor effect and the main source of force fluctuations comes from the neighbors, i.e., the microstructure. Furthermore, the contributions from large spheres are consistently larger than those from small spheres, which is also valid in the comparison between $l-s$ and $s-l$ interactions. Both of these two observations can be applied to $F_L$ and $T_{\perp}$ as shown in Figure \ref{bi_inter_FL} and \ref{bi_inter_TL} respectively, where the fore-aft symmetry only exists at $\Re=1$ and then the critical regions drift correspondingly toward the $1-4$, and $2-3$ quadrants when $\Re$ increases. This trend coincides with the probability distribution maps as well. In addition, the upstream neighbors exhibit more effect on $F_x$ and less effect on $F_L$ at higher $\Re$, and vice versa, whereas the influential ranges of $T_L$ are considerably smaller than those of forces, thus fewer neighbors are required in torque predictions. 

Finally, we carry out a similar analysis for the second nearest neighbor and display its contribution to the total force/torque exerted on a target sphere (solid black circle) by fixing the nearest neighbor (hollow black circle) at an upstream position with a gap $=0.2$ (only for demonstration purpose). Again, the outer white spaces enclosed by the black dashed circles define regions where the neighbor cannot reside, and the outermost green dashed circles reflect the range of influential neighbors, i.e., the position of the 50th and 20th neighbor in force and torque predictions respectively. We can identify eight distinct modes for trinary interactions in bidisperse particle-laden flows, and denote each mode with a format of target sphere $-$ nearest and second nearest neighbors. We select the case $\left(\Re,\phi,d_l^*/d_s^*,V_l^*/V_s^*\right)=\left(100,15\%,2,2\right)$ as an example here, and similar observations can be made in any other case: (\expandafter{\romannumeral1}) Contributions of the second nearest neighbor are constantly smaller than those of the nearest neighbor as shown by the legend of each subplot, which is the result of prominently larger weight given to the nearest neighbor to address its dominant role. (\expandafter{\romannumeral2}) Similar to the previous observation in binary interactions, large spheres always produce a greater influence regardless of the interaction mode, implying that our PINN model places particular emphasis on the size of neighboring spheres throughout the learning process. (\expandafter{\romannumeral3}) The contour shapes of all three contributions $F_x$ $F_L$ and $T_{\perp}$ of the second nearest neighbor are identical to those of the nearest neighbor and independent of the interaction mode. This is not surprising as in the design of our PINN architecture, each input neighbor is separately sent to and processed by the same NN block, thus generating the same influential pattern with only different magnitude. This unified function/contribution representation is the core feature of the pairwise interaction assumption. It simplifies the multi-body interaction problem and enables to separate the influence of a specific neighbor from that of the other neighbors and to evaluate it based on the properties of each pair of spheres, whereas the order of spheres in the local assembly only affects the weights of superposition but not the general pattern.

\begin{figure}
	\centering
	\captionsetup[subfigure]{oneside,margin={0.3cm,0.cm}}
	\begin{subfigure}[c]{0.9\linewidth}		
		\resizebox{\linewidth}{!}{\includegraphics{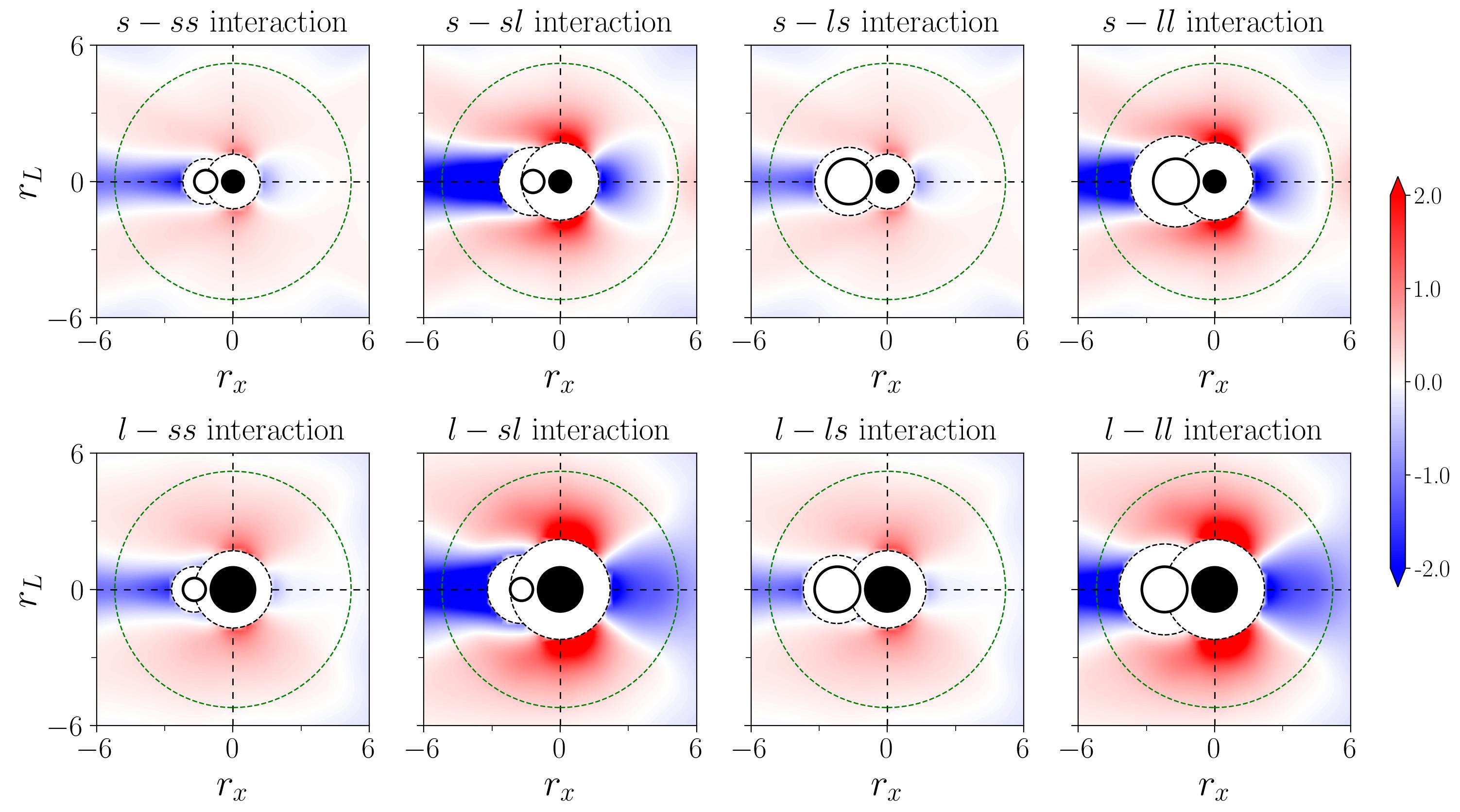}}
		\vspace{-0.7cm}
		\subcaption{$F_x$}
		\vspace{0.cm}
		\label{tri_inter_Fx}
	\end{subfigure}
	\begin{subfigure}[c]{0.9\linewidth}		
		\resizebox{\linewidth}{!}{\includegraphics{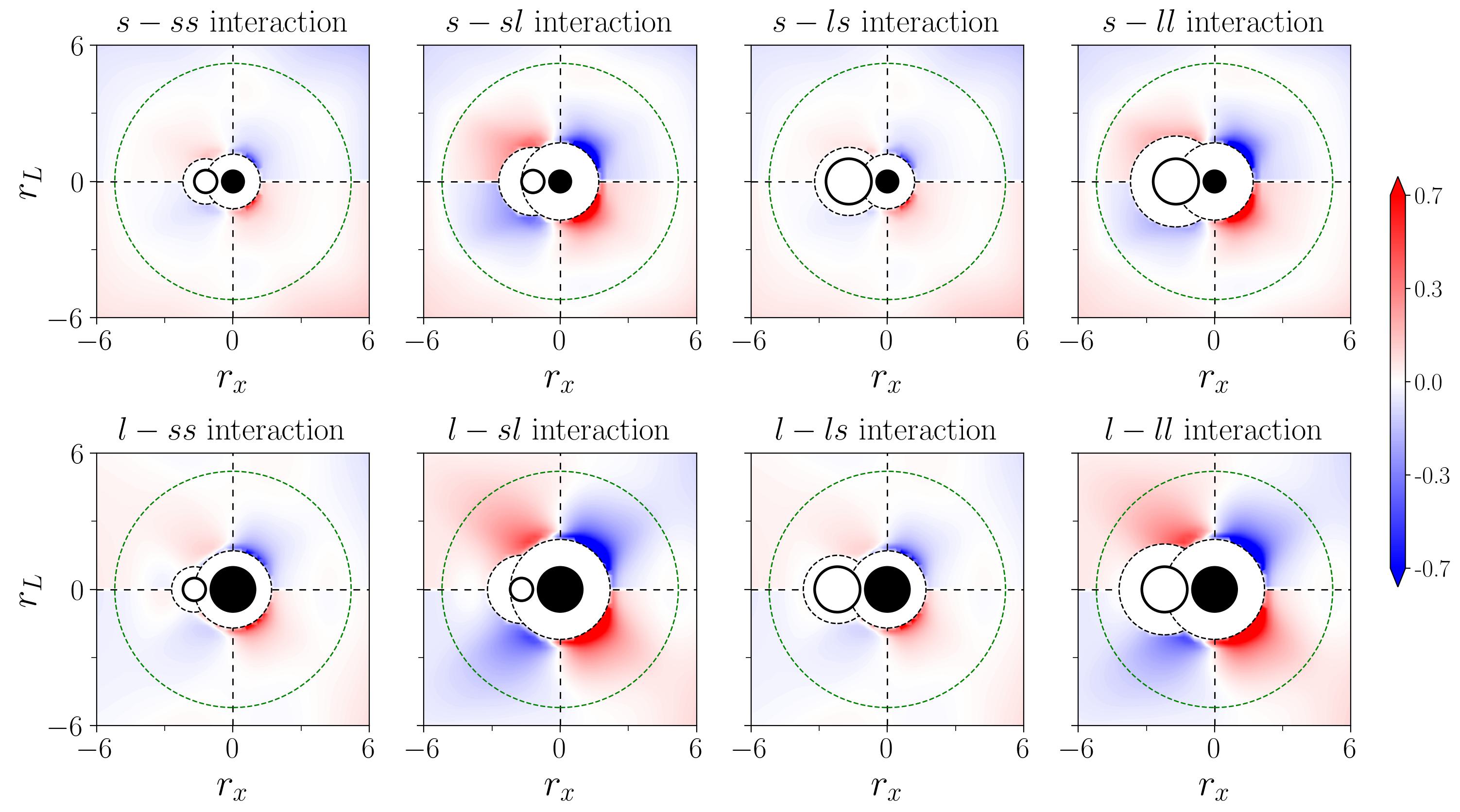}}
		\vspace{-0.7cm}
		\subcaption{$F_L$}
		\label{tri_inter_FL}
	\end{subfigure}
	\begin{subfigure}[c]{0.9\linewidth}		
		\resizebox{\linewidth}{!}{\includegraphics{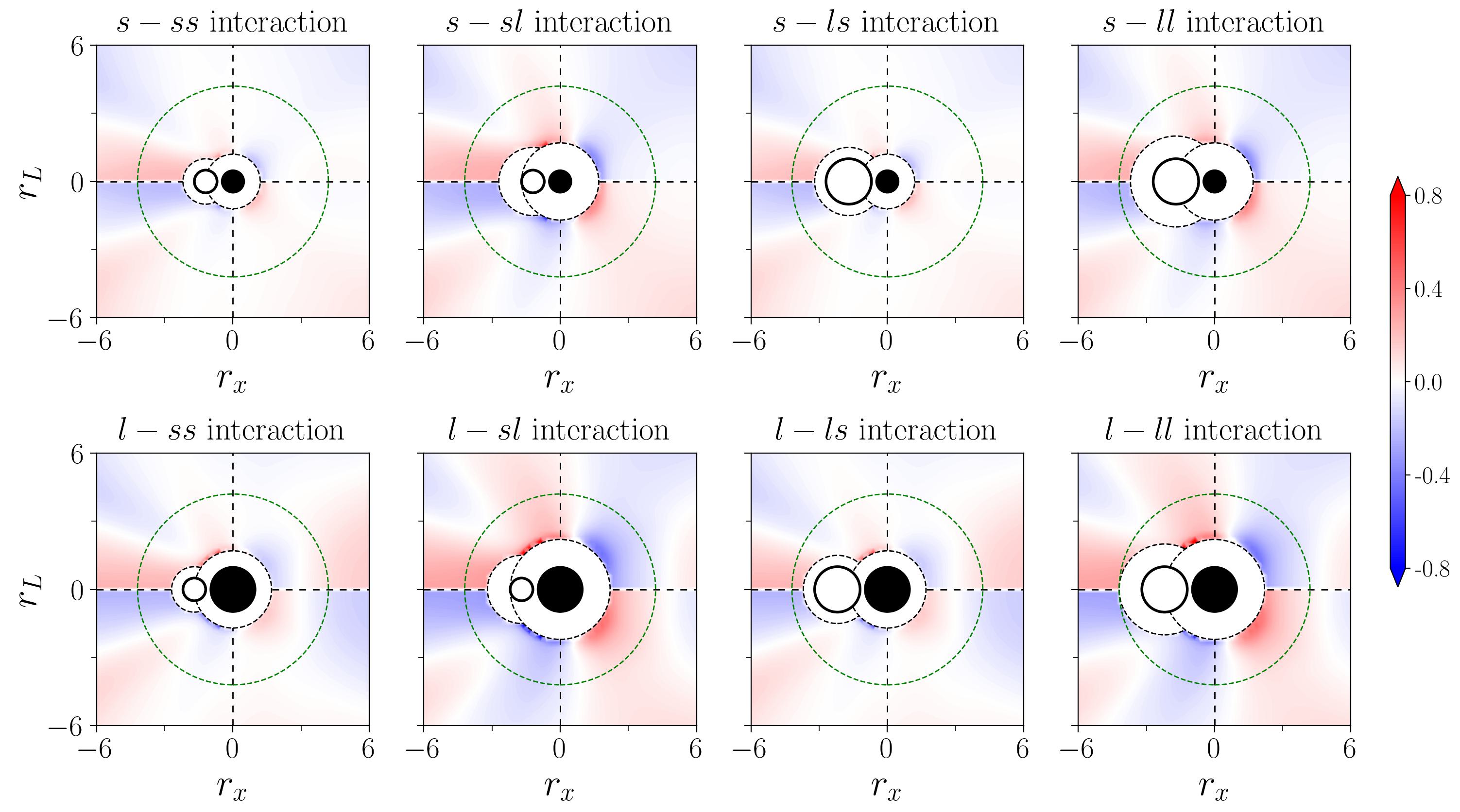}}
		\vspace{-0.7cm}
		\subcaption{$T_{\perp}$}
		\label{tri_inter_TL}
	\end{subfigure}
	\caption{Contribution of the second closest neighbor to the reference particle with its closest neighbor fixed at upstream position in different interaction modes.}
	\label{tri_inter}
\end{figure}


\section{Conclusions \label{sec-con}}

Physics-informed ML and data-driven approaches have been widely adopted to tackle fluid mechanics problems and several works have demonstrated their promising application to predict the hydrodynamic force/torque fluctuations in monodisperse particle-laden flow problems. These fluctuations  were usually neglected in the past although PR-DNS data validated that the standard deviation of drag distribution has a comparable magnitude as its mean value \cite{Akiki2016,SeyedAhmadi2020}. However, we still lack the knowledge of force/torque distributions in polydisperse particle-laden flows due to the increased computing cost of the corresponding PR-DNS, thus compromising further research including the extension of ML techniques to this field. Consequently, we performed a series of PR-DNS of the flow past a random array of stationary bidisperse and polydisperse spheres in our previous work \cite{Cheng2023}. Results exhibited that the force and torque exerted on each sphere class still follow a Gaussian distribution with a non-negligible span, and the probability distribution maps of different binary interaction modes in bidisperse particle-laden flows are strikingly similar in terms of contour shapes and critical regions. On top of these numerical findings, we are now able to extract the key characteristics from this large dataset and leverage a PINN model to exploit the statistical relationship between the force/torque fluctuations and relevant variables in bidisperse particle-laden flows as a starting point.

Inspired by several example works on monodisperse particle-laden flows \cite{Moore2019,SeyedAhmadi2022,Siddani2023}, we formulated our problem of interest based on the same pairwise interaction assumption \cite{Akiki2017a,Akiki2017b} that approximates the hydrodynamic force/torque of each sphere in a stationary random array as a linear superposition of individual contributions from $M$ influential closest neighbors. As another core assumption, the similar probability distribution maps of different interaction modes imply a unified function to describe such a force/torque contribution from each neighbor regardless of its size or distance to its target sphere. Therefore, we construct a compact PINN architecture that takes relative position and diameter $\left\{\br_j,d_j\right\}$ of neighbor $j$ as inputs to evaluate its effect on a target sphere through the same NN block, which is a reflection of unified function representation similar to that proposed in \cite{SeyedAhmadi2022}. To incorporate the effects of local mesoscale flow and microstructure, we select the local average velocity, local solid volume fraction and target sphere diameter $\left\{\bRe_i,\phi_i,d_i\right\}$ as additional inputs and process them with another NN block. The scalar outputs of two NN blocks experience an element-wise multiplication to manifest the contribution of each neighbor with the consideration of local flow conditions. At the final stage, we compute the total force/torque exerted on a target sphere using weighted sums according to the pairwise interaction assumption.

We determined the number of influential neighbors as $M=50$ and $20$ in force and torque predictions respectively, and correspondingly the depth (number of layers) and width of both NN blocks being $\left(3,30\right)$ and $\left(3,20\right)$ according to the grid-search results. We trained our PINN model with PR-DNS datasets of eight PR-DNS cases spanning $1\leq\Re\leq100$, $10\%\leq\phi\leq40\%$, $1.5\leq d_l^*/d_s^*\leq2.5$ and $1\leq V_l^*/V_s^*\leq 4$. Through a comparison of PINN model predictions and PR-DNS data, we demonstrated great overall performance with $R^2$ in a range of $0.706\sim0.892$ for $F_x$, $0.669\sim0.871$ for $F_L$ and $0.741\sim0.910$ for $T_{\perp}$ predictions. The success of the current model does not only reflect in the close performance of small and larger spheres with distinct force/torque distributions, but also in the minor difference in $R^2$ of training and test sets which indicates excellent model generalization to the data that it has not trained on. We further compared the force/torque distribution statistics of PINN predictions to those of PR-DNS data and obtained very similar distributions with both approaches, including a good agreement of their respective standard deviations and a remarkably small error of $0.18\%$ on the mean $F_x$ even in the worst case scenario. Subsequently, we repeated the training procedures with the same hyper-parameters on multiple combined datasets to construct a universal predictive model, and evaluated the model performance on the unseen data again generated by PR-DNS with new sets of input parameters. Satisfactory predictions were reported in most cases, indicating good interpolation capability of our universal PINN model within the considered range of fluid and particle properties. Finally, we showed the contribution maps of the nearest and second nearest neighbors in different binary and trinary interaction modes. We noticed that the critical regions, i.e., positive or negative force/torque contributions, are qualitatively identical to those shown in our previous probability distribution maps, implying great interpretability provided by our PINN predictions.

As we mentioned previously, the bidisperse particle-laden flow cases examined here are the starting point of our "proof-of-concept" study. Our ultimate goal is to evolve from a problem-specific model to a versatile model which is capable of coping with other particle-laden flow problems involving more complicated configurations, e.g., flows laden with polydisperse spheres and/or non-spherical particles. This cannot be fulfilled without the prior knowledge we supplied in the present work and the fine-tuned design of the corresponding PINN architecture. As pointed out by Seyed-Ahmadi and Wachs \cite{SeyedAhmadi2022},  the simple fully-connected NN model even fails to deliver correct predictions with an increasing number of influential neighbors. On the contrary, the PINN architecture used in the present work possesses several key advantages that will potentially facilitate its extension to more complex particle-laden flow problems:
\begin{enumerate}
	\item Due to the common NN block shared by different neighbors, we are theoretically able to incorporate an infinite number of influential neighbors as inputs without adjusting the overall PINN architecture, whereas the size (both depth and width) of hidden layers must be accordingly expanded in this case for a fully-connected NN model. This is crucial for strongly bidisperse and polydisperse particle-laden flows as the number of influential neighbors should be increased when we consider a large sphere diameter ratio. (as an example, $M=50$ and 20 in our bidisperse cases with only $1.5\leq d_l^*/d_s^*\leq2.5$, while $M\leq30$ and 10 in monodisperse cases in previous works). 
	\item The unified function representation \cite{SeyedAhmadi2022} (similar to the order invariance assumption in \cite{Moore2019}) has been proven to be suitable in bidisperse cases based on our observations of probability distribution maps obtained with different binary interaction modes, hence inspiring this particular architecture of shared NN block for every neighbor-related input sets. Therefore, we assume that this feature could be applied to any flow laden with polydisperse spherical particles with the same inputs $\left\{\br_j,d_j\right\}$ of neighbor $j$ (still needs to be trained and validated with PR-DNS data), and similarly to flows laden with non-spherical particles with additional independent variables to characterize neighbors, e.g., sphericity and aspect ratio, to be inserted in the approximating contribution functions and incorporated into the input space.
	\item We separately evaluated the effect of local mesoscale flow and neighborhood based on $\left\{\bRe_i,\phi_i,d_i\right\}$ with another NN block, and combined this effect with each neighbor contribution throughout an element-wise multiplication as adopted in \cite{Siddani2023}. Such an operation is more physically reasonable than directly supplying these inputs to the linear superposition which seems oversimplified in \cite{SeyedAhmadi2022}. Thanks to this second NN block, we are also allowed to consider many other intermediate variables, e.g., pressure components as suggested in \cite{Muralidhar2019}, to improve the force/torque predictions.
\end{enumerate}

It is worth highlighting another bonus promoted by such a compact PINN architecture that is also pointed out in the previous section: the total number of unknown parameters is tremendously decreased compared to a fully connected NN. For example, the number of unknown parameters is 4100/4150 in force predictions ($F_L$ has two components) and 1940 in torque predictions, whereas it is stunningly 25850 for a (6,50) NN and 81500 for a (5,100) NN as reported in \cite{Siddani2023} (the reason is that such a fully connected NN is assigned to correlate a much larger number of input variables, i.e., inputs of all neighbors versus inputs of a single neighbor in our PINN model). Consequently, our PINN model is much more efficient not only in terms of training (i.e., optimizing the value of unknown parameters), but more importantly in terms of evaluating the force/torque distributions for a random sphere assembly. In addition, the elu activation function used in our model has a simpler form than the hyperbolic tangent unit that also contributes to faster computation. As an example, it takes $<1$ second for our model to evaluate the force or torque component in an array of $4000$ spheres with 16 CPU-core parallel computing. Therefore, it is promising to implement our PINN model in EL simulations to provide the missing hydrodynamic particle force and torque exerted by the locally complex fluid flow, which are usually oversimplified by any average drag closure (as the other components have a mean value of $0$). Undoubtedly, our PINN model is still slower than the average drag closure computation since these closures typically have a concise functional form dependent on mesoscale variables only, such as $\Re$, $\phi$ and $d_i^*$. Eventually, there is no free lunch but there always exists an appropriate trade-off between accuracy and computing cost, depending on the level of fidelity expected from a simulation. We believe that our PINN model that provides good performance in both aspects can arguably become a strong alternative to enhance the EL simulations.

We would like to share final remarks about the fundamental assumption of linear superposition for force/torque exerted on a targe sphere. In our work, we consistently apply this linear superposition and achieve great predictions, most probably because the upper bound of the selected range $1\leq\Re\leq\ 100$ is relatively low, whereas we can reasonably anticipate that a non-linear relation should be employed at higher $\Re$. Siddani and Balachandar \cite{Siddani2023} suggest a hierarchical interpretation of hydrodynamic force/torque and expand it into unary, binary, trinary and higher-order interaction terms. This is undoubtedly adequate but training a NN model to learn trinary and higher order interactions is far more challenging. However, if we are able to find the relation between binary and trinary terms, and successively between $N$th and $\left(N+1\right)$th terms, a non-linear relation between force/torque and the binary interaction terms only might be equivalently deduced. Correspondingly in the PINN architecture, we merely need to add an extra NN block to the output layers to approximate this non-linear function, where each binary interaction contribution is taken as input and force/torque as output. We leave this to our future research.

\clearpage
\section*{References}

\bibliographystyle{elsarticle-num}
\bibliography{biblio}

\newpage

\listoffigures

\end{document}